\documentclass[useAMS,usenatbib,twocolumn,superscriptaddress,reprint,showpacs]{revtex4}
\usepackage{times} 
\usepackage{amsmath}
\usepackage{rotating}
\usepackage{amssymb}

\newcommand{\hMsol}{{\>h^{-1}\rm M}_\odot}

\begin{document}

\title{The cosmic X-ray and gamma-ray background from dark matter annihilation}
\author{Jes\'us Zavala}
\thanks{CITA National Fellow}
\email{jzavalaf@uwaterloo.ca}
\affiliation{Max-Planck-Institut f\"{u}r Astrophysik, Karl-Schwarzschild-Stra\ss{}e 1, 85740 Garching
bei M\"{u}nchen, Germany}
\affiliation{Department of Physics and Astronomy, University of Waterloo, Waterloo, Ontario, N2L 3G1, Canada\footnote{Current affiliation}}
\author{Mark Vogelsberger}
\affiliation{Harvard-Smithsonian Center for Astrophysics, 60 Garden Street, Cambridge, MA 02138, USA}
\author{Tracy R. Slatyer}
\affiliation{School of Natural Sciences, Institute for Advanced Study, Princeton, NJ 08540, USA}
\author{Abraham Loeb}
\affiliation{Harvard-Smithsonian Center for Astrophysics, 60 Garden Street, Cambridge, MA 02138, USA}
\author{Volker Springel}
\affiliation{Heidelberg Institute for Theoretical Studies, Schloss-Wolfsbrunnenweg 35, 69118 Heidelberg, Germany}
\affiliation{Zentrum f\"ur Astronomie der Universit\"at Heidelberg, Astronomisches Recheninstitut,
  M\"{o}nchhofstr. 12-14, 69120 Heidelberg, Germany}

%\date{Accepted ???. Received ???; in original form ???}

%\pagerange{\pageref{firstpage}--\pageref{lastpage}} \pubyear{2009}

%\maketitle

%\label{firstpage}

\begin{abstract}
The extragalactic background light (EBL) observed at multiple
wavelengths is a promising tool to probe the nature of dark
matter. This radiation might contain a significant contribution from
gamma-rays produced promptly by dark matter particle annihilation in
the many halos and subhalos within our past-light
cone. Additionally, the electrons and positrons produced in the
annihilation give energy to the cosmic microwave photons to populate
the EBL with X-rays and gamma-rays. To study these signals, we create
full-sky maps of the expected radiation from both of these
contributions using the high-resolution Millennium-II simulation of
cosmic structure formation. Our method also accounts for a possible
enhancement of the annihilation rate by a Sommerfeld mechanism due to
a Yukawa interaction between the dark matter particles prior to
annihilation. We use upper limits on the contributions of unknown
sources to the EBL to constrain the intrinsic properties of dark
matter using a model-independent approach that can be employed as a
template to test different particle physics models. These upper limits
are based on observational measurements spanning eight orders of
magnitude in energy (from soft X-rays measured by the CHANDRA
satellite to gamma-rays measured by the {\it Fermi} satellite), and on
expectations for the contributions from non-blazar active galactic
nuclei, blazars and star forming galaxies. To exemplify this approach,
we analyze a set of benchmark Sommerfeld-enhanced models that give the correct abundance of dark 
matter, satisfy constraints from the cosmic microwave background, and fit the
cosmic ray spectra measured by PAMELA and {\it Fermi} without any contribution from local substructure. 
We find that these models are in conflict with the EBL constraints unless the contribution of unresolved 
substructure is small and the dark matter annihilation signal dominates the EBL. We conclude that provided the 
collisionless cold dark matter paradigm is accurate, even for conservative estimates of the contribution from 
unresolved substructure and astrophysical backgrounds, the EBL is \emph{at least} as sensitive a probe 
of these types of scenarios as the cosmic microwave background. More generally, our results disfavor an
explanation of the positron excess measured by the PAMELA satellite
based only on dark matter annihilation in the smooth Galactic dark matter halo.
\end{abstract}


\maketitle

\section{Introduction}

A broad class of particles known as Weakly Interacting Massive Particles (WIMPs),
are the best studied and arguably the most
favored candidates to be the primary component of cosmic dark matter. The most
prominent example of such particles is the neutralino that arises naturally in 
supersymmetry (SUSY); for recent reviews on neutralino dark matter
  see \cite{2009EPJC...59..557S,2010ARA&A..48..495F}. WIMPs
can explain the
observed abundance of dark matter in a natural way and  because
they behave as cold dark matter (CDM) they are also favored by the
prevailing $\Lambda$CDM cosmology, which is the most successful model
of structure formation to date. Furthermore, most WIMPs are
particularly appealing because they offer a relatively high chance of
detection in the near future, through: i) direct detection experiments
on Earth looking for the recoil of ordinary matter by scattering of
WIMPs, and ii) indirect searches that look for standard model
particles produced in the annihilation of WIMPs.

A number of observations in recent years have highlighted anomalies
that might be caused by dark matter annihilation. The excess of
positrons in cosmic rays above 10~GeV reported by the PAMELA
experiment \cite{2009Natur.458..607A} is one of these observations,
and although other astrophysical sources, such as pulsars
\cite{Hooper-Blasi-DarioSerpico-09} and supernova remnants
\cite{Fujita-09}, could explain the signal, the possibility of dark
matter annihilation remains attractive and has motivated a significant
number of papers on the topic. It is however necessary to invoke large
annihilation rates and specific annihilation channels to explain the
anomalies with dark matter annihilation alone
\cite{Bergstrom-Edsjo-Zaharijas-09}. These rates are orders of
magnitude larger than the ones obtained assuming the standard values
for the annihilation cross section that give the correct relic density
of dark matter. Due to their higher densities, substructures in the
local dark matter distribution can boost the annihilation rates, but
not to the required level \cite{Lavalle-08}. The non-linear collapse
of collisionless dark matter halos leads to the formation of
caustics, which due to their high density could significantly increase
the annihilation rate. However, this increase is actually much less significant than previously
thought \cite{2009MNRAS.400.2174V,2010arXiv1002.3162V}. In the inner parts of halos, it is essentially negligible
and can not be invoked to explain the high annihilation rates required
to explain the PAMELA measurements.

Alternatively, an elegant solution may lie in an enhancement of the
annihilation cross section by a Sommerfeld mechanism produced by the
mutual interaction between WIMPs prior to their annihilation
\cite{2004PhRvL..92c1303H,2009PhRvD..79a5014A,2009PhRvD..79h3523L}. This
enhancement could easily be large enough to explain the anomalous
excess of cosmic ray positrons.

The annihilation rate can however not be arbitrarily large either as
it is constrained by different observables. For example, dark matter
annihilation can ionize and heat the photon-baryon plasma at
recombination, creating perturbations in the Cosmic Microwave
Background (CMB) angular power spectrum
\cite{Galli-09,Slatyer-Padmanabhan-Finkbeiner-09,2011arXiv1103.2766H}. It can also alter
the relic abundance of dark matter significantly
\cite{Dent-Dutta-Scherrer-09,2010PhRvD..81h3502Z}, and produce
important 
$\mu-$ and $y-$type distortions of the CMB
\cite{2010PhRvD..81h3502Z,2011JCAP...01..016H}.
All these observables hence constrain the
degree to which the Sommerfeld mechanism can enhance the cross
section. Nevertheless, it is possible to satisfy all these constraints
and at the same time explain the positron excess measured by PAMELA \cite{Finkbeiner:2010sm}.

An additional set of observations with the potential to constrain the
annihilation cross section can come from the analysis of the
extragalactic background radiation at multiple wavelengths. The
annihilation of WIMPs can manifest itself as a cosmic background
radiation with gamma-ray photons being produced promptly in all
extragalactic sources with high dark matter density
\cite{Ullio-02,2003MNRAS.339..505T,2005PhRvL..94q1302E,Ando-Komatsu-06,Cuoco-07,deBoer-07,Cuoco-08,Fornasa-09,2009PhRvL.102x1301S}. This
gamma-ray radiation is complemented towards lower energies by a
diffuse extragalactic background in photons that were not produced
directly in the annihilation but gained energy via inverse Compton
scattering off the energetic electrons and positrons produced during
the annihilation
\cite{2009JCAP...07..020P,2010PhRvD..81d3505B,2010JCAP...07..008H}.

The data collected by several telescopes over the last decades have
given us a measurement of the extragalactic radiation background from
soft X-rays to hard gamma-rays (e.g. \cite{2008ApJ...689..666A}). In
this broad energy range, most of the radiation is expected to be
produced by astrophysical mechanisms different from dark matter
annihilation. This has been partially confirmed by accounting for the
radiation of known sources and by estimating the contribution of an
expected population of sources yet to be observed. This combined set
of observations and expectations puts strong constraints on the
contribution of dark matter annihilation, being specially stringent in
the soft-X-ray regime where $\sim90\%$ of the emission comes from
X-ray point sources, mostly Active Galactic Nuclei (AGN)
\cite{2007ApJ...661L.117H}, and on the gamma-ray regime where blazars
and star-forming galaxies are expected to contribute significantly to
the background radiation, at the level of $\sim70\%$
\cite{2010ApJ...720..435A,2009MNRAS.400.2122A}.

The hypothetical background radiation coming from (or being
up-scattered in) all dark matter halos and their subhalos within our
past light cone has been studied by different authors in the past
using analytic approaches to model cosmic structure formation. An
approach based directly on high-resolution numerical simulations is
however desirable since it more accurately captures the non-linear
phase of the evolution, even though the simulation imposes a
resolution limit for smallest structure.  It is then possible to
construct simulated sky-maps of the background radiation that give a
more complete description of the signal. Such an approach was
developed in \cite{2010MNRAS.tmp..453Z} to analyze the extragalactic
gamma-ray radiation produced in situ by annihilation using the
state-of-the-art Millennium II simulation
\cite{2009MNRAS.398.1150B}. In this paper, we extend this approach to
include the contribution from CMB photons scattered by the electrons
and positrons produced during annihilation.

By using this approach we are also able to easily include a
velocity-dependent annihilation cross section via a Sommerfeld
mechanism. Typically, the enhancement is inversely proportional to the
local velocity dispersion of dark matter particles. Since our method
is based on average values of the annihilation rate inside halos and
their subhalos, the Sommerfeld enhancement is simply given by the
mean velocity dispersion in each halo (subhalo), which is available in
the simulation and can be measured accurately.

The paper is organized as follows. In Section~\ref{formalism}, we
outline the formalism to calculate the extragalactic background
radiation coming from in situ and up-scattered photons. A description
of the Sommerfeld enhancement model we used and its implementation is
given in Section~\ref{s_enh}. The observational upper limits and main
results of our work on the cosmic background radiation are presented
in Section~\ref{results}. Finally we present a summary and conclusions
in Section~\ref{concl}.

\section{Annihilation radiation formalism}\label{formalism}

Our goal is to analyze the cosmic dark matter annihilation background
(CDMAB), or more specifically, the radiation produced by dark matter
annihilation in all extragalactic sources integrated over all
redshifts along the line-of-sight of a fiducial observer, located at
$z=0$, for all directions on its two-dimensional full sky.  To this
end, we first define the local photon emissivity:
\begin{equation}\label{emiss}
\mathcal{E}=\frac{f_{\rm WIMP}}{2} E \rho_{\chi}({\vec x})^2
S(\sigma_{\rm vel}({\vec x})), \quad f_{\rm WIMP}= \frac{{\rm d}N}{{\rm
    d}E}\frac{\langle\sigma v\rangle_0}{m^2_{\chi}}
\end{equation}
where $m_{\chi}$ and $\rho_{\chi}$ are the mass and density of
WIMPs, $\langle\sigma v\rangle_0$ is the thermally averaged product of the
constant s-wave annihilation cross section and the velocity in the absence of Sommerfeld
enhancement, and ${\rm d}N/{\rm d}E$ is the differential photon yield per annihilation. 
The velocity dispersion dependent factor
$S(\sigma_{\rm vel})$ boosts the value of $\langle\sigma v\rangle_0$ through a Sommerfeld 
mechanism (see section~\ref{s_enh}). We note that for consistency, the value of 
$\langle\sigma v\rangle_0$ should also give the correct dark matter relic 
density.

The CDMAB is given by the specific intensity, the
energy of photons received per unit area, time, solid angle and energy
range:
\begin{equation}\label{intensity}
I=\frac{1}{4\pi}\int
\mathcal{E}(E_{0}(1+z),z)\frac{{\rm d}r}{(1+z)^4}e^{-\tau(E_{0},z)},
\end{equation}
where the integral is over the whole line of sight, $r$ is the
comoving distance and $E_{0}$ is the photon energy measured by the
observer at $z=0$. Note that $\mathcal{E}$ is evaluated at the
blue-shifted energy $(1+z)E_{0}$ along the line-of-sight to compensate
for the cosmological redshifting. The exponential term with an
effective optical depth $\tau(E_{0},z)$ takes into account the
absorption of photons by the matter and radiation field along the
line-of-sight. The relevant processes of photon absorption and their
treatment are described in Appendix B.

In this work, we focus on two different contributions to the
differential photon yield per annihilation event ${\rm d}N/{\rm
  d}E$. In the following, we describe these contributions that we refer
to as \textit{in situ} and \textit{up-scattered} photons.

\subsection{In situ photons}\label{in_situ}

The in-situ photons are directly created due to the annihilation
process. They are in the gamma-ray energy range and are produced by
three mechanisms: (i) continuum emission following the decay
of neutral pions produced during the hadronization of the primary
annihilation products; (ii) monoenergetic lines for
WIMP annihilation in two-body final states containing photons;
(iii) internal bremsstrahlung when the final products of annihilation
are charged, leading to the emission of an additional photon in the
final state. Process (i) is dominant at most gamma-ray energies, but
processes (ii) and (iii) produce distinctive spectral features
intrinsic to the annihilation phenomenon. This in situ contribution to
the CDMAB has been studied in detail before by different authors. In
this work we follow the analysis of \cite{2010MNRAS.tmp..453Z},
extending their results to lower energies as described below.

\subsection{Up-scattered photons}\label{up_scat}

The up-scattered photons originate in differently produced background
photons that gain energy due to their interactions with particles
produced in the annihilation of dark matter. We concentrate
exclusively on Inverse Compton (IC) scattering as the mechanism
contributing to the up-scattering of these photons, and on the CMB as
the main photon background.  There are additional backgrounds, like
stellar and infrared light, that are dominant close to galactic discs 
in the center of relatively massive halos. However, most of the CDMAB 
comes from the integrated effect of low mass halos and subhalos (see Appendix \ref{astro_factor}).
In these places, the stellar component is rather small and the mean number density of
starlight and infrared photons is much lower than that of the CMB.

Electrons and positrons are the annihilation byproducts participating
in the scattering. These particles are quite energetic and have
therefore usually a large $\gamma=1/\sqrt{1-(v/c)^2}$ factor. This
implies that they can up-scatter low energy photons to significantly
higher energies, because of the $\gamma^2$-dependence of the peak
energy of up-scattered photons. In this process, CMB photons increase
their energy into the X-ray and low gamma-ray regimes
\cite{2009JCAP...07..020P}.

The differential electron (and positron) yield that is relevant for
the IC up-scattering of the CMB photons is found by solving a
diffusion equation that takes into account the diffusion and energy
losses of these particles:
\begin{equation}\label{diffusion}
\frac{\partial}{\partial t}\frac{{\rm d}n_{\rm e}}{{\rm d}E_{\rm e}} =
\nabla \left[ D_{\rm e} \nabla\frac{{\rm d}n_{\rm e}}{{\rm d}E_{\rm
      e}}\right] + \frac{\partial}{\partial E_{\rm e}} \left[ b_{\rm
    e} \frac{{\rm d}n_{\rm e}}{{\rm d}E_{\rm e}}\right] + Q_{\rm e},
\end{equation}
where ${\rm d}n_{\rm e}/{\rm d}E_{\rm e}$ is the equilibrium electron
spectrum, $D_{\rm e}=D_{\rm e}(E_{\rm e},{\vec x})$ is the diffusion
coefficient, $b_{\rm e}=b_{\rm e}(E_{\rm e},{\vec x})$ is the energy
loss term and $Q_{\rm e}=Q_{\rm e}(E_{\rm e},{\vec x})=\mathcal{E}_{\rm e}/E_{\rm e}$\footnote{Here $\mathcal{E}_{\rm e}$ is the local electron (and positron) emissivity, 
analogous to the photon emissivity defined in Eq.~(\ref{emiss}).} 
is the source function. Spatial diffusion due to scattering on the inhomogeneities
of the ambient magnetic field can be neglected. This is because
spatial diffusion is only relevant at relatively small scales, within
a few kpc of the center of dark matter halos
\cite{2006A&A...455...21C}. However, we are interested in a
cosmological background radiation where most of the signal in a given
area in the sky comes from unresolved sources far away, where spatial
diffusion is clearly irrelevant. In this case, the steady-state solution to Eq.~(\ref{diffusion})
can be approximated by:
\begin{eqnarray}\label{eq:eqdist}
\frac{{\rm d}n_{\rm e}}{{\rm d}E}(E_{\rm e},z)&\approx&\frac{1}{b_{\rm
    e}(E_{\rm e},z)}\int_{E_{\rm e}}^{m_\chi} {\rm d}E_{\rm
  e}^\prime\ Q_{\rm e}(E_{\rm e},{\vec x})\nonumber\\
&=&\frac{\langle\sigma v\rangle_0}{2}\left(\frac{\rho_{\chi}({\vec x})}{m_{\chi}}\right)^2
S(\sigma_{\rm vel}({\vec x})) \frac{{\rm d}{\tilde n}_{\rm e}}{{\rm d}E}(E_{\rm e},z). \ \ \ \ \ \
\end{eqnarray}

The energy loss rate, $b_{\rm e}(E_{\rm e},z)$, for electrons and
positrons receives contributions from different interaction processes:
IC scattering with ambient photons, synchrotron radiation in the
ambient magnetic field, Coulomb scattering with free electrons,
ionization of atoms and bremsstrahlung radiation in interactions with
the ambient matter field. As we explain in Appendix A, among all these
cooling processes we only consider the first one since it dominates
the photon energy range we are ultimately interested in. The energy
loss term in Eq.~(\ref{eq:eqdist}) is hence given by Eq.~(\ref{loss_IC}).

We further assume that the electrons and positrons produced in the
annihilation process lose energy and reach equilibrium
instantaneously (in a cosmological time frame), scattering the CMB
photons at the same redshift at which the annihilation takes
place (e.g. \cite{2005PhRvD..72b3508P}). These up-scattered photons have a differential photon spectrum
given by:
\begin{equation}
\frac{{\rm d}N_{\rm IC}}{{\rm d}E}(E,z)= \int {\rm d}E_{\rm
  e}\ \frac{{\rm d}{\tilde n}_{\rm e}}{{\rm d}E}(E_{\rm e},z)\ {\tilde P}_{\rm
  IC}\left(E,E_{\rm e},z\right),
\end{equation}
where the IC power per scattered photon energy is: 
\begin{equation}
{\tilde P}_{\rm IC}\left(E,E_{\rm e},z\right)=c \int {\rm d}{\tilde E}\ n_{\rm
  CMB}({\tilde E},z)\sigma_{\rm KN}(E,E_{\rm e},{\tilde E}).
\end{equation}
Here, $n_{\rm CMB}({\tilde E},z){\rm d}{\tilde E}$ is the number density of CMB photons 
in the energy range $({\tilde E}, {\tilde E}+{\rm d}{\tilde E})$
at redshift $z$:
\begin{equation}
n_{\rm CMB}(\tilde E,z){\rm d}{\tilde E}= \frac{8\pi}{(h c)^3} \frac{{\tilde
    E}^2{\rm d}{\tilde E}}{\exp[{\tilde E}/(k_{\rm B}~T_0(1+z))]-1},
\end{equation}
where $T_0=2.725$~K is the CMB
temperature today, which increases with redshift like
$(1+z)T_0$. Finally, $\sigma_{\rm KN}$ is the differential
Klein-Nishina cross-section for IC scattering
\begin{equation}
\sigma_{\rm KN}(E,E_{\rm e},{\tilde E}) = \frac{3 \sigma_{\rm T}}{4
  \tilde{E}} \left(\frac{m_{\rm e} c^2}{E_{\rm e}}\right )^2
\ G\left(q,\Gamma_{\rm e}\right),
\end{equation}
where $\sigma_{\rm T}$ is the Thomson cross-section, $m_{\rm e}$ the electron mass, and
\begin{equation}
G\left(q,\Gamma_{\rm e}\right) = \left[ 2 q \ln q + (1+2 q)(1-q) +
  \frac{\left(\Gamma_{\rm e} q \right)^2 (1-q)}{2 \left( 1+
    \Gamma_{\rm e} q\right)}\right], 
\end{equation}
with
\begin{equation}
\Gamma_{\rm e}= \frac{4 \tilde E E_{\rm e}}{(m_{\rm e}c^2)^2}, \quad q
= \frac{E}{\Gamma_{\rm e} \left( E_{\rm e} - E \right)}.
\end{equation}
Note that the Klein-Nishina cross-section depends on the increased
energy of the up-scattered photon $E$, the energy of the original CMB
photon $\tilde E$, and on the energy $E_{\rm e}$ of the electron that
does the IC scatter. The limits for the various integrals above are
given by the kinematic constraint of the IC scattering requiring
$1/[4(E_{\rm e}/(m_{\rm e}c^2))^2]<q<1$.

\section{Sommerfeld enhancement}\label{s_enh}

We include in our analysis a scenario where the annihilation is
enhanced by the Sommerfeld mechanism
(e.g. \cite{2004PhRvL..92c1303H,2005PhRvD..71f3528H,2009PhRvD..79a5014A,2009PhRvD..79h3523L}),
restricting it to the case where the interaction between WIMPs
prior to annihilation is mediated by a scalar boson of mass $m_\phi$
through a Yukawa potential with coupling constant $\alpha_{\rm c}$
(e.g. \cite{2009PhRvD..79h3539B}).  This case encompasses the large
majority of the models that are typically used in the literature to
account for the Sommerfeld enhancement, including the most common of
these where $S\propto1/\sigma_{\rm vel}$ (the so-called ``$1/v$''
boost). Even in models with nearly-degenerate interacting states and/or multiple 
force carriers, while the details of the enhancement differ, the general features 
remain similar. The enhancement saturates at sufficiently low
velocities due to the finite range of the Yukawa interaction. For
certain combinations of $\alpha_{\rm c}$ and $m_\phi$, resonances associated
with zero-energy bound states appear \footnote{ The related phenomenon of 
radiative capture into WIMPonium, allowed for $m_\phi < \alpha_{\rm c}^2 m_\chi/4$, 
can increase the effective enhancement factor substantially for small $m_\phi$ 
\cite{2009PhLB..676..133M, 2009PhRvD..79e5022S}.}. Close to these
resonances, the enhancement gets significantly larger for low
velocities and scales as $1/\sigma_{\rm vel}^2$. In this case, the
enhancement also saturates eventually due to the finite lifetime of
the states. Regardless of the values of the parameters, the boost to
the cross-section disappears for velocities comparable to the speed of
light. This argument has often been invoked to infer that the dark
matter relic density is unaffected by the Sommerfeld enhancement, but
it has been shown recently that this assumption is not
correct \cite{2010PhRvD..81h3502Z}. 

A detailed description of the Sommerfeld model studied here has been
presented elsewhere
(e.g. \cite{2009PhRvD..79h3539B,2009PhRvD..79h3523L}). For the
purposes of this work, we follow the description of
\cite{2010PhRvD..81h3502Z} and mention that the enhancement $S(\sigma_\mathrm{vel})$ 
to the $s$-wave contribution to the annihilation rate is given by:
\begin{align}\label{thermal}
&\langle\sigma v\rangle=\langle\sigma v\rangle_0 S(\sigma_{\rm vel}),
  \nonumber \\ &S(\sigma_{\rm vel}) = \left(\frac{1}{2\sigma_{\rm
      vel}^3 \sqrt{\pi}}
  \int_0^1S(\beta)\beta^2e^{-\beta^2/4\sigma_{\rm vel}^2}~{\rm d}\beta\right),
\end{align}
where $\beta=v_{{\rm rel}}/c$ is the relative velocity between the annihilating
pair\footnote{Unless otherwise stated, velocities are given in units
  of the speed of light~$c$.}.

For definiteness, we choose two sets of parameters that fall within
currently favored regions of the parameter space
(e.g. \cite{2009PhRvD..79h3539B}): case i) off-resonance:
$m_\phi/m_\chi=5\times10^{-4}$, $\alpha_{\rm c}=3\times10^{-2}$ and case ii)
near-resonance:
$m_\phi/m_\chi=2.98\times10^{-4}$, $\alpha_{\rm c}=3\times10^{-2}$. The
former is representative of the standard ``$1/v$'' boost with a maximum
enhancement $S_{\rm max}\sim2000$ for $\sigma_{\rm
  vel,max}\sim10^{-5}$. The latter is a typical resonance case with
$S\propto1/\sigma_{\rm vel}$ at intermediate velocities and
$S\propto1/\sigma_{\rm vel}^2$ at low velocities up to a saturation
$S_{\rm max}\sim10^6$ for $\sigma_{\rm vel,max}\sim6\times10^{-7}$.

By solving the Schr\"odinger equation for s-wave annihilation in the
non-relativistic limit, we obtain $S(\beta)$ for the two cases chosen
above, and use Eq.~(\ref{thermal}) to get the average annihilation
boost $S(\sigma_{\rm vel})$ for each halo.  Since we can estimate the
change on the values of $S_{\rm max}$ and $\sigma_{\rm vel,max}$ for a
different set of parameters, the results we obtain later using these
representative cases serve us to analyze the whole range of
possibilities that are expected for a Sommerfeld mechanism produced
by a Yukawa potential.

The Sommerfeld enhancement alters the relic density of dark matter \cite{Dent-Dutta-Scherrer-09,2010PhRvD..81h3502Z}. 
During freeze-out, while the 
Sommerfeld enhancement is generally $\mathcal{O}$(1), it is not negligible and 
can consequently have an $\mathcal{O}$(1) effect on the relic density (requiring 
a reduction of the underlying annihilation cross section to compensate). After kinetic 
decoupling of the dark matter from the radiation bath, the typical velocities of the 
dark matter particles decrease rapidly: even for non-resonant (but unsaturated) enhancement,
the enhanced annihilation rate keeps pace with the universe's expansion, and for resonant 
enhancement the dark matter annihilations can actually \emph{recouple} (depending on the 
relative temperatures of freeze-out, kinetic decoupling and saturation of the enhancement), 
greatly reducing the relic density. Once the enhancement saturates, the annihilation rate 
no longer keeps pace with the expansion rate, and the comoving density of dark matter 
remains fixed. In all cases there is a significant effect on the relic density, and in order to
produce the correct abundance today, the value of the annihilation
cross-section before the onset of the enhancement needs to be smaller
than for the case without Sommerfeld enhancement.

This result is relevant because it implies that any particle physics
model without enhancement chosen to satisfy the observational bounds
on the abundance of dark matter needs to be revised once the
enhancement is included to test whether or not it still gives the
correct relic density. The fully consistent way to do so is to
incorporate the Sommerfeld enhancement into a Boltzmann code and
re-sample the parameter space of that particular model to find allowed
regions. Here, we follow a simpler approach. According to
\cite{2010PhRvD..81h3502Z}, the value of $\langle\sigma
v\rangle_0$ should be lower by a
factor between 1 and 10 compared to the case without enhancement in order to get the correct
relic density. The
precise reduction factor, $f_{\Omega}$, depends on the intensity of
the enhancement: $f_{\Omega}\sim0.1$ near resonances and
$f_{\Omega}\sim0.5$ off-resonance.  Therefore, by multiplying
$\langle\sigma v\rangle_0$ by the corresponding $f_{\Omega}$ factor,
we roughly take into account the effect on the relic density. In this
way, a model without enhancement that gives the correct relic density
with $\langle\sigma v\rangle_0$ will also give the right relic
abundance with Sommerfeld enhancement, provided its annihilation cross
section in the early Universe is chosen to be $f_{\Omega}\langle\sigma
v\rangle_0$.

\section{Extragalactic CDMAB}\label{results}

The procedure we follow to construct the simulated sky maps of the
contribution of dark matter annihilation to the X-ray and gamma-ray
extragalactic background radiation is essentially an extension of the
one discussed in \cite{2010MNRAS.tmp..453Z}. For a detailed
description of the map-making technique we used, we hence refer the
reader to section 5.1 of that paper.

According to Eq.~(\ref{emiss}), the local annihilation rate depends on
the square of the local density of dark matter. For the computation of the
cosmological background from an N-body simulation,
it is more reliable to use
analytically integrated quantities over whole dark matter halos
(based on scaling laws tested with extremely high-resolution
  simulations of MW-like halos \cite{2008Natur.456...73S}) instead of
trying to use individual simulation particles directly, which are
subject to stronger resolution effects and numerical noise \cite{2010MNRAS.tmp..453Z}.
Using
this method, each pixel in our sky maps receives contributions of all
intervening resolved halos and subhalos along its corresponding past
light cone. Additionally, we add the expected contribution of
unresolved structures down to the damping scale limit of WIMPs
($10^{-6}$M$_{\odot}$, see sections 5.2-5.4 of \cite{2010MNRAS.tmp..453Z}). Since we are
exploring the case with Sommerfeld enhancement in this paper, the
formulae given in \cite{2010MNRAS.tmp..453Z} need to be altered
accordingly. In Appendix~C we describe how we accomplish this.

The signal depends of course on the value of the photon yield ${\rm
  d}N/{\rm d}E$ as well, which contains contributions from in situ and
up-scattered photons. These are determined by the intrinsic properties
of WIMPs. As an example, we take a neutralino with a main
annihilation channel into $b{\bar b}$. In particular, we use a
benchmark point within the minimal supergravity (mSUGRA) framework
(model L in Table~I of \cite{2010MNRAS.tmp..453Z}). This benchmark
point has $m_{\chi}\sim185$~GeV with annihilation into $b{\bar b}$
with a $99\%$ branching ratio, and $\langle\sigma
v\rangle\sim6.2\times10^{-27}{\rm cm}^3{\rm s}^{-1}$. It belongs to
the so-called ``bulk region'' within the mSUGRA 5-dimensional
parameter space that is consistent with current constraints on the
relic density of neutralinos (if neutralinos make up for all the
observationally inferred dark matter density). We obtain the photon,
electron and positron yields for this model using the numerical code
{\small DarkSUSY} \cite{2004JCAP...07..008G,2005NewAR..49..149G} with
the interface {\small ISAJET} \cite{2003hep.ph...12045B}.

\begin{figure}
\center{
\includegraphics[height=8.0cm,width=8.0cm]{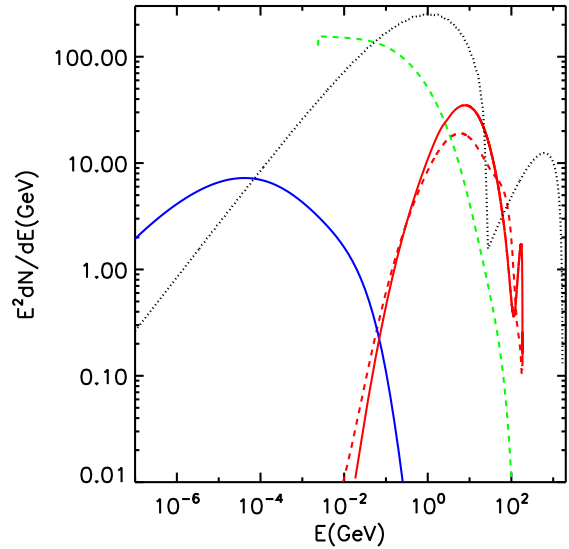}
}
\caption{Photon yield for dark matter annihilation in the X-ray and
  gamma-ray energy range for a $\sim185$~GeV neutralino annihilating
  into $b{\bar b}$.  The contributions from in-situ and up-scattered
  CMB photons are shown with solid red and blue lines,
  respectively. For reference, the in-situ and equilibrium electron
  yields from annihilation are shown with dashed red and green lines
  respectively (the equilibrium spectrum as defined in
  Eq.~(\ref{eq:eqdist}) was scaled by a factor of $10^{-16}$ to show
  it in the same figure). Also shown in the figure with a black
    dotted line is the total photon yield from benchmark model 1 of Table 1, 
    see section \ref{bench}.}
\label{fig:spectra} 
\end{figure}

In Fig.~\ref{fig:spectra}, we show the final photon yield spectrum for
dark matter annihilation in the X-ray and gamma-ray energy range for
the example just described. The contributions from in situ and
up-scattered CMB photons are shown with solid red and blue lines,
respectively. For reference, the in situ and equilibrium electron
(positron) yields from annihilation are shown with dashed red and
green lines, respectively \footnote{The equilibrium spectrum as defined
  in Eq.~(\ref{eq:eqdist}) was scaled by a factor of $10^{-16}$ to
  show it in the same figure.}. The main bump and secondary peak that
are clearly shown for the in situ photons correspond to the two main
mechanisms mentioned in section \ref{in_situ}, neutral pion decay and
internal bremsstrahlung, respectively. The figure shows clearly that
although in this case the largest photon yield is in the gamma-ray
regime, there is a significant amount of X-ray radiation produced by
IC scatter of the CMB photons. The contribution from up-scattered
CMB photons is the dominant feature for other particle physics models. As 
an example of this we show in Fig.\ref{fig:spectra} the total photon yield for 
one of a set of benchmark models that we use later in section \ref{bench}. The shape 
and normalization for this benchmark model 1 (see Table 1) are representative
of all the benchmark models we will use.

\begin{figure}
\centering
\includegraphics[height=8.0cm,width=8.0cm]{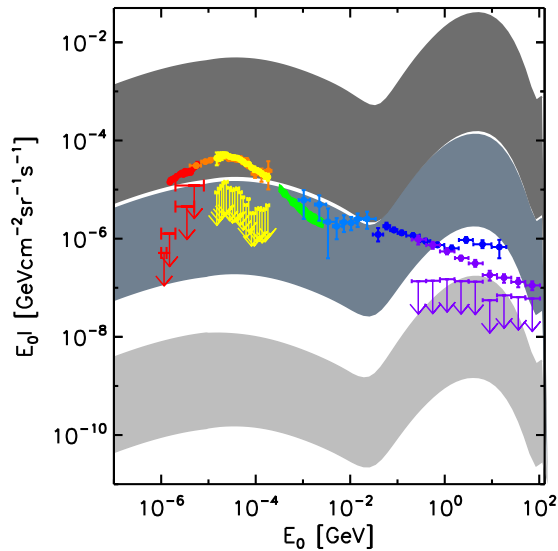}
\caption{CDMAB spectrum including in-situ and up-scattered CMB photons
  (gamma-rays from annihilation and CMB photons up-scattered to X-ray
  energies by electrons and positrons of the annihilation) for the
  cases of no-enhancement (light-gray), enhancement away from a
  resonance (medium-gray) and near a resonance (dark-gray). The upper
  and lower limits of each stripe bracket the uncertainty in the
  extrapolation of unresolved subhalos in the simulation. All cases
  are for a model with annihilation mainly into $b{\bar b}$,
  $m_{\chi}\sim185$~GeV and $\langle\sigma
  v\rangle_0\sim6.2\times10^{-27}{\rm cm}^3{\rm s}^{-1}$. They all
  give approximately the correct relic density.  Observations from
  soft X-rays to gamma-rays are marked with red to violet, following
  approximately a rainbow color pattern: red symbols
  \cite{2009A&A...493..501M}, red arrows
  (Chandra, \cite{2007ApJ...661L.117H}), orange symbols
  (INTEGRAL, \cite{2007A&A...467..529C}), yellow symbols
  (SWIFT BAT, \cite{2008ApJ...689..666A}), yellow arrows
  \cite{2009ApJ...696..110T}, green area
  (SMM, \cite{1997AIPC..410.1223W}), light blue
  (COMPTEL, \cite{2000AIPC..510..467W}), blue
  (EGRET, \cite{2004ApJ...613..956S}), violet
  ({\it Fermi-LAT}, \cite{2010JCAP...04..014A}), violet arrows
  \cite{2010ApJ...720..435A,2009MNRAS.400.2122A}.  The points with
  error bars are absolute measurements with $2\sigma$ or $1\sigma$
  errors.  The arrows pointing downwards are best estimate upper limits
  of the unresolved component of the signal, that is, the signal that
  can not be accounted for by already known or expected sources.}
\label{fig:EBL_Somm}
\end{figure} 

Fig.~\ref{fig:EBL_Somm} shows the contribution from dark matter
annihilation to the X-ray and gamma-ray extragalactic background
radiation for the particular SUSY model described above. The
case without Sommerfeld enhancement is shown within the light-gray
shaded region.  This region is bracketed by the maximum and minimum
values of the extrapolation for unresolved subhalos, which
encompasses the astrophysical uncertainties in the contribution by
low-mass subhalos that can not be resolved by the Millennium II
simulation (see Appendix \ref{astro_factor}). The medium-gray and
dark-gray shaded regions are for the cases with Sommerfeld
enhancement, off- and near-resonance, respectively.

According to Fig.~\ref{fig:EBL_Somm}, any model with a photon yield
similar to the one we used as an example (see Fig.~\ref{fig:spectra}),
and with a mass $\sim200$~GeV and $\langle\sigma
v_0\rangle\sim6\times10^{-27}{\rm cm}^3{\rm s}^{-1}$, could be ruled
out, depending on how relevant the contribution from
unresolved halos and subhalos is.  Any significant Sommerfeld
enhancement is clearly ruled out by absolute measurements of the
background in this case.

It is important to note that the Sommerfeld mechanism is of relevance
for neutralinos only for masses $\gtrsim$TeV in the case of a minimal SUSY
model like mSUGRA
(e.g. \cite{2005PhRvD..71f3528H,2009PhRvD..79h3523L}). In this case,
the force carriers responsible for the enhancement are the W and Z
gauge bosons \footnote{Although the formalism required to compute the
  enhancement is more complicated than the one given in section
  \ref{s_enh}, which is strictly valid for scalar boson carriers, the
  results are qualitatively similar
  \cite{2009PhRvD..79h3523L}.}. Therefore, boosts of order $\sim1000$
or even larger are only possible for neutralinos with much higher
masses than the model we have chosen as an example in
Figs. \ref{fig:spectra} and \ref{fig:EBL_Somm}. The net effect of a
higher neutralino mass in the input photon and positron (electron)
spectra is a shift of the X-ray and gamma-ray peaks shown in these
figures towards higher energies. Nevertheless, the ${\rm d}N/{\rm d}E$
spectrum shown in Fig.~\ref{fig:spectra} is generic for any model with
a WIMP annihilating mainly into $b{\bar b}$. If such a generic
model allows the inclusion of a new scalar boson responsible of the
Sommerfeld enhancement, then the formalism described in
Section~\ref{s_enh} is applicable and Fig.~\ref{fig:EBL_Somm} shows
the expected level of enhancement of the CDMAB due to this mechanism.

The symbols shown in Fig.~\ref{fig:EBL_Somm} represent inferences for
the extragalactic X-ray and gamma-ray background radiation based on
observational data as described in the following.

\subsection{Observations}\label{obs}

We are interested in measurements of the cosmic background radiation
in an energy range going from soft X-rays to gamma-rays: $0.1~{\rm
  keV}\leq E_{0}\lesssim100~{\rm GeV}$. Because for $E_{0}<1$~keV the
signal is completely dominated by galactic and local emission that
varies with time and position, estimates of the extragalactic emission
at these energies have not been possible
\cite{2006ApJ...645...95H,2007ApJ...661L.117H}.  In the range $1~{\rm
  keV}\leq E_{0}\leq200~{\rm keV}$, the extragalactic X-ray background
has been studied in detail by satellites such as CHANDRA, SWIFT and
INTEGRAL. We take the absolute measurements obtained using the latter
two satellites according to the analysis of
\cite{2009A&A...493..501M} (red symbols in Fig.~\ref{fig:EBL_Somm}),
\cite{2007A&A...467..529C} (orange symbols) and
\cite{2008ApJ...689..666A} (yellow symbols). At intermediate
energies, $300~{\rm keV}\leq E_{0}\leq30~{\rm MeV}$, the measurements
come from the Solar Maximum Mission (SMM) \cite{1997AIPC..410.1223W}
and COMPTEL \cite{2000AIPC..510..467W}. These measurements are shown
with green and light blue points, respectively. Finally, observations
based on EGRET \cite{2004ApJ...613..956S} and recently on {\it Fermi}
\cite{2010JCAP...04..014A} have estimated the cosmic background in
gamma-rays from $40~{\rm MeV}$ to $~100$~GeV. These estimates are
shown with dark blue and purple symbols, respectively.

The observational data we have described above give a measurement of
the total extragalactic X-ray and gamma-ray background radiation. Over
the full energy range, most of the signal is expected to come from
photons produced by different astrophysical sources in mechanisms that
are unrelated to dark matter annihilation. The contribution from the
latter is likely to be a subdominant component of the total signal,
which is especially true at lower
energies. Upper limits to the
emission that have not been accounted for by known sources for
$E_{0}<8$~keV, have been found using CHANDRA data (red arrows) \cite{2007ApJ...661L.117H}. 
Approximately less than $10\%$
of the integrated specific intensity
is unresolved  between $1$~keV$<E_{0}<8$~keV. For hard X-rays
($10~{\rm keV}\leq E_{0}\leq200~{\rm keV}$), most of the emission is
expected to come from Compton-thin Active Galactic Nuclei (AGN). We
use the model presented in \cite{2009ApJ...696..110T} to put a
conservative upper limit on the unresolved component of the emission
at these energies (yellow arrows)\footnote{Specifically, we take the
  obscured and unobscured Compton-thin AGN contributions to the signal
  in this energy range (red and blue solid lines in Fig. 5
    of \cite{2009ApJ...696..110T}). We do not include the contribution
  from Compton-thick AGN that has a more uncertain modeling. We note
  that due to this, upper limits to the contribution of sources other
  than AGN in this energy range are expected to be lower than those
  shown in Fig.~\ref{fig:EBL_Somm}.}. The modeling in the MeV range is
more uncertain. According to some analyses, blazars are thought to
contribute significantly to the radiation \cite{2009ApJ...699..603A}, but
others argue for non-blazar AGNs as the main
contributors to the MeV radiation \cite{2010arXiv1001.0103I}. We will not attempt to model the
contribution of these sources due to this controversy, but we note
that the constraints on the contribution of dark matter annihilation
to the MeV background are expected to be significantly lower than
those seen in Fig.~\ref{fig:EBL_Somm}. Recently, the {\it Fermi-LAT} 
collaboration has estimated the blazar contribution to the gamma-ray background 
in the $0.1-100$~GeV energy range. Its total specific intensity in this 
energy range (i.e. its integrated flux between these energies) down to the
minimum detected source flux is $\sim16\%$ of the derived value for the 
cosmic gamma-ray background \cite{2010ApJ...720..435A}. We use the energy spectrum 
given by these authors (see their Table~6 and Fig.~20) to account for the blazar 
contribution noting that this is a conservative estimate since undetected sources
certainly contribute to the signal, see below. Star forming galaxies are also expected to be
a significant source of gamma-rays in this energy range. We use the
model by \cite{2009MNRAS.400.2122A} to include this contribution (see their Fig.~1), which
accounts for $\sim53\%$ of the total specific intensity.
We note that the energy spectrum of the contribution of star forming galaxies to the cosmic
gamma-ray background (as plotted in Fig.~\ref{fig:EBL_Somm}, i.e. $E_0I$) peaks
at $E_0\sim0.3$~GeV, dominating over the blazar spectrum, and drops more steeply towards 
higher energies than the total background. At $E_0\gtrsim10$~GeV blazars 
dominate over star forming galaxies with a spectrum shallower than the observed background. In this 
way, both populations combined account for $\sim86\%$ ($\sim46\%$) of the measured specific 
intensity at $E_0\sim0.3$~GeV ($E_0\sim70$~GeV). Over the whole $0.1-100$~GeV energy range, 
they account for $\sim69\%$ of the total integrated flux.
Based on this, the corresponding upper limits 
on the contribution from additional sources are shown with violet arrows in
Fig.~\ref{fig:EBL_Somm}.

We should comment on the uncertainties associated to
the contribution of blazars and star forming galaxies. For the latter, these 
are connected to the gamma-ray luminosity function of galaxies which is
ultimately related to the time-dependent global star formation rate density. 
The possible behaviors of the gamma-ray
luminosity function and the most relevant sources of uncertainty 
in the model (more importantly the cosmic star formation rate and the normalization
given by the inferred gamma-ray luminosity of the Milky-Way) have been considered
by \cite{2010ApJ...722L.199F} using a similar modeling to that of \cite{2009MNRAS.400.2122A}. 
The authors find that 
star forming galaxies account for between $10\%$ to $90\%$ of the EBL 
measured by FERMI at $E_0\sim0.3$~GeV 
(see Fig.~1 of \cite{2010ApJ...722L.199F}) with a spectral shape very similar to the model 
we have chosen here. Since the contribution from star forming galaxies is quite uncertain and
since the fiducial model we use lies closer to the upper value of this contribution, we 
explore below the effects that a lower contribution has in our 
results (see section \ref{bench}). If the 
observed count distribution of blazars is extrapolated to zero flux then their
contribution to the total observed signal between $0.1$~GeV and $100$~GeV is
$\sim23(\pm9)\%$ (including statistical and systematic uncertainties) \cite{2010ApJ...720..435A}. 
As we mentioned before,
this percentage drops to $\sim16(\pm9)\%$ when only sources down to the minimum flux are
considered. At $E_0\sim0.3$~GeV the minimum contribution from blazars to the
measured specific intensity is $\sim7\%$ (including uncertainties). Taking the 
lower limits of all these uncertainties into account, 
star forming galaxies and blazars would contribute minimally by $\sim17\%$ at 
$E_0\sim0.3$~GeV, a factor of 5 lower than the estimate we
use here.

\subsection{Constraints on particle physics models}

With the procedure we have previously outlined, we can compare the
prediction of any given particle physics model with the observational
upper limits shown in Fig.~\ref{fig:EBL_Somm}.  The model gives a
photon and a positron (electron) input spectra from the annihilation, and
our map-making code produces a simulated map for a prescribed
energy. A full spectrum can be then produced once maps at different
energies are constructed.

It is possible however to present robust limits on the part of the
signal that only depends on the intrinsic properties of WIMPs,
namely $f_{\rm WIMP}$ in Eq.~(\ref{emiss}). This can be done by
noting that there exists a redshift $z^{\ast}$ along the line-of-sight
for which Eq.~(\ref{intensity}) can be written as:
\begin{eqnarray}\label{intensity_2}
I(E_{0})=\frac{c}{8\pi}E_{0}f_{{\rm WIMP}}(E_{0}(1+z^{\ast}))\nonumber\\ \int\frac{\rho_{\chi}^2({\vec
    x},z)S(\sigma_{{\rm vel}}({\vec
    x},z))}{(1+z)^3}\frac{e^{-\tau(E_{0},z)}}{H(z)} {\rm d}z,
\end{eqnarray}
where $H(z)$ is the Hubble parameter \footnote{Here we have
  changed the comoving distance $r$ in Eq.~(\ref{intensity})
  for the redshift $z$ and use the mean value theorem of integral
  calculus, which can be used because the remainder
  integrand in Eq.~(\ref{intensity_2}) is always positive in the
  interval of integration.}. In general, we do not know the value of
$z^{\ast}$, it is model dependent. Nevertheless, we can safely
approximate the upper limit of the integral in Eq.~(\ref{intensity_2})
by $z=4$ in the case of the lowest X-ray energies and by $z=1$ for the
higher gamma-ray energies (and values in between for intermediate
energies). This is because $\sim90\%$ of the signal is produced for
$z<4$ ($z<1$) in the former (latter) case. The relevant redshift range
is significantly smaller at higher energies where photon absorption
plays an important role.  This approximation is good enough for any
model with a photon yield spectrum ${\rm d}N/{\rm d}E$ similar to the
one depicted in Fig.~\ref{fig:spectra}.  
More generally, for any model with a photon yield that is monotonically 
decreasing with energy, the large majority of the signal at any given energy 
would come from relatively low redshifts, since the contribution from higher 
redshifts would correspond to higher-energy (and hence less abundant) initial
photons. This is true because the astrophyiscal part of the specific intensity 
that goes in the integrand of Eq.~(\ref{intensity_2}), excluding the absorption factor, is 
essentially flat with redshift (see for example Fig. 1 of \cite{2010JCAP...04..014A}).
Thus, we can use the observed upper limits on the unaccounted contribution
to $I(E_{0})$, and the values of this same quantity predicted by a
reference particle physics model to estimate upper limits on $f_{\rm
  WIMP}(E_{0}(1+z^{\ast}))$:
\begin{equation}\label{fsusy_obs}
f_{\rm WIMP}(E_{0}(1+z^{\ast}))\leq f_{\rm WIMP}^{\rm
  REF}(E_{0}(1+z^{\rm REF}))\frac{I^{\rm OBS} (E_{0})}{I^{\rm
    REF}(E_{0})},
\end{equation}
where the values associated with the reference model are given with a
superscript ${\rm REF}$. The values of $z^{\ast}$ and $z^{\rm REF}$
are in the interval $(0,4)$ for $E_{0}\sim10^{-5}$~GeV and in the
interval $(0,1)$ for $E_{0}\sim10$~GeV. 
By choosing $z_{\rm REF}=0$, so $f_{\rm WIMP}^{\rm REF}$ is evaluated at the 
measured energy rather than the (higher) effective energy of injection, we obtain 
a conservative 
constraint, since this function is monotonically decreasing with respect to energy. 
Taking a higher value for $z_{\rm REF}$ will strengthen the bound. By choosing $z^{\ast}=0$, 
we evaluate $f_{\rm WIMP}$ at the lowest possible energy for the purpose of comparing 
to the limit: this is not conservative for models where $f_{\rm WIMP}$ is a monotonically 
falling function of energy, in the sense that taking a larger $z^{\ast}$ leads to a weaker 
limit, but since the signal is dominated by the lowest redshifts, the resulting 
uncertainty is quite small. We denote the upper bound on $f_{\rm WIMP}(E)$ obtained by 
setting $z^{\ast}=z_{\rm REF}=0$ by $f_{\rm WIMP}^{\rm MAX}(E)$.

\begin{figure}
\centering
\includegraphics[height=8.0cm,width=8.0cm]{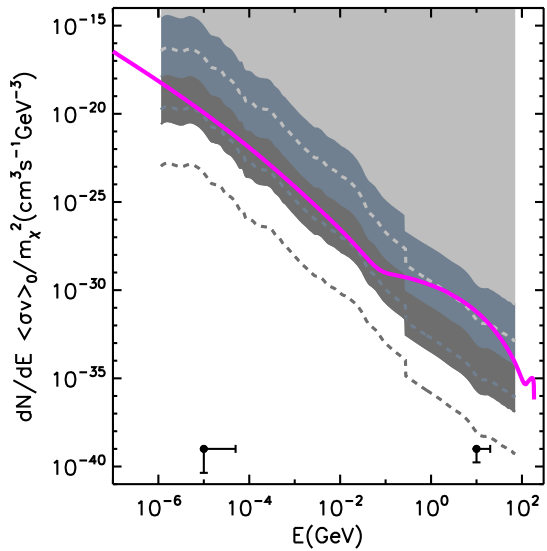}
\caption{Limits on the value of $f_{\rm WIMP}$ for dark matter
  annihilation according to observations of the cosmic X-ray and
  gamma-ray background radiation. The light-gray, medium-gray and
  dark-gray areas mark exclusion regions for the case with no
  Sommerfeld enhancement, off- and near-resonance enhancement,
  respectively, in the case where the contribution of unresolved
  substructures to the signal is minimal. The dashed lines show how
  these regions are extended if this contribution is maximal. The area
  between the limit of each shaded region and its corresponding dashed
  line encompasses the uncertainty on the contribution of unresolved
  subhalos. The symbols with error bars in the bottom show the
  theoretical uncertainty on the construction of this figure from
  Fig.~\ref{fig:EBL_Somm}; see text for details. The thick magenta
  line is for the SUSY model we have used as an example: a
  $\sim185$~GeV neutralino annihilating into $b{\bar b}$ with
  $\langle\sigma v\rangle_0=6\times10^{-27}$cm$^3$s$^{-1}$.}
\label{fig:fSUSY_bounds}
\end{figure} 

We take as a reference model the example we have used throughout the
text, and compute $f_{\rm WIMP}^{\rm MAX}$ for the cases without
enhancement, and with Sommerfeld boost, off-resonance and
near-resonance. The exclusion regions we obtain are shown in
Fig.~\ref{fig:fSUSY_bounds} with the light-gray, medium-gray and
dark-gray regions, respectively, for these three cases, assuming  a
minimum extrapolation for the contribution of unresolved substructures
to the simulated maps. The dashed lines show how these exclusion
limits are extended if a maximum extrapolation is taken.  The right-
and down-wards error bars in the figure mark the uncertainty in the
values of $z^{\ast}$ and $z^{\rm REF}$, respectively. As mentioned before,
the amplitude of the uncertainty depends on the value of the observed
energy, being lower for higher energies. The thick magenta line shows
the value of $f_{\rm WIMP}$ for our reference case.

\begin{figure}
\centering
\includegraphics[height=8.0cm,width=8.0cm]{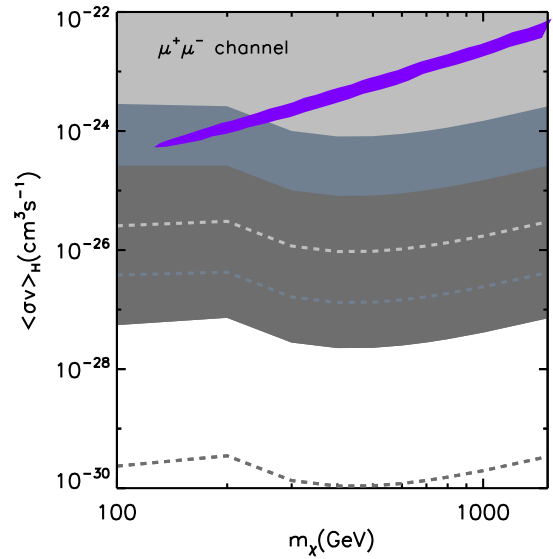}
\caption{Constraints on the local value of the thermally averaged
  annihilation cross section (assuming a MB velocity distribution with
  $\sigma_{\rm vel}=150\, {\rm km\, s}^{-1}$) as a function of WIMP mass
  for annihilation into $\mu^{+}\mu^{-}$ final states. These
  constraints come from observations of the cosmic X-ray and gamma-ray
  background radiation.  The violet contour shows the $2\sigma$ best
  fit region of this model to the PAMELA positron data as presented in
  \cite{2010JCAP...04..014A}.  The other line styles and colors are
  as in Fig.~\ref{fig:fSUSY_bounds}.}
\label{fig:cross_bounds}
\end{figure} 

The validity of a given model can be tested directly using
Fig.~\ref{fig:fSUSY_bounds} without the need of computing the CDMAB for
this model. Keep in mind that for the cases with Sommerfeld
enhancement, the value of $\langle\sigma v\rangle_0$ in $f_{\rm WIMP}$ is
the value of the s-wave annihilation cross section without Sommerfeld enhancement.

Once a specific model is chosen, ${\rm d}N/{\rm d}E$ is calculated and
Fig.~\ref{fig:fSUSY_bounds} can be used to produce constraints on
$\langle\sigma v\rangle_0$ as a function of WIMP mass. As an
example, we take a model with annihilation into leptons, specifically
$\mu^{+}\mu^{-}$ with a branching ratio of $100\%$. Models such as
this are typically used in the literature to explain the anomalous
abundance of positrons in cosmic rays above 10~GeV reported by the
PAMELA satellite \cite{Bergstrom-Edsjo-Zaharijas-09}. We use {\small
  DarkSUSY} \cite{2004JCAP...07..008G} to compute the in situ
positron (electron) and photon spectra. The resulting total photon yield has a very similar
shape to the black dotted line shown in Fig.~\ref{fig:spectra}, which is by the way shared by all
the benchmark models we describe in section \ref{bench}. For this case, and for
$m_{\chi}>100$~GeV, the up-scattered photons contribute dominantly to
the background radiation. Instead of showing constraints on
$\langle\sigma v\rangle_0$ we show in Fig.~\ref{fig:cross_bounds} the
constraints on $\langle\sigma v\rangle_{\rm H}=S(\sigma_{\rm vel}=150{\rm
  kms^{-1}})\langle\sigma v\rangle_0$, which is a thermal average over
a Maxwell-Boltzmann (MB) velocity distribution with a velocity
dispersion of $150~{\rm km\, s^{-1}}$($5\times10^{-4}~c$). This
roughly corresponds to the estimated local dark matter one-dimensional velocity
dispersion. For both cases of Sommerfeld enhancement we are
considering: $S(\sigma_{\rm vel}=150\,{\rm km\, s^{-1}})\sim230$. With
this choice, we can compare the constraints coming from the
extragalactic background radiation with the local values of
$\langle\sigma v\rangle$ that better fit the PAMELA data for an
explanation of the positron excess based solely on dark matter
annihilation. The violet contour in Fig.~\ref{fig:cross_bounds} shows
the $2\sigma$ best fit region according to
\cite{2010JCAP...04..014A}. For this particular model with
annihilation into $\mu^{+}\mu^{-}$ with a branching ratio of $100\%$,
the constraints we find do not favor an explanation of the PAMELA
data based only on dark matter annihilation for $m_{\chi}>260$~GeV
(a similar conclusion was found
  in \cite{2009JCAP...07..020P,2010JCAP...04..014A}).  A large saturated
Sommerfeld enhancement ($S_{\rm max}>2000$) essentially rules out this
possibility.

We note that any model tested using Fig.~\ref{fig:fSUSY_bounds} also
needs to be checked for consistency with the correct relic
density. Contrary to Fig.~\ref{fig:EBL_Somm} that was used to
exemplify a case where a specific model gives the correct dark matter
abundance with and without enhancement (recall that for the former we
multiplied the value of $\langle\sigma v\rangle$ without enhancement
by a factor $f_{\Omega}$ to accomplish this, see end of section
\ref{s_enh}), the upper limit on $f_{\rm WIMP}$ in Eq.~(\ref{fsusy_obs})
deliberately does not take this into account. In this way, the upper
limits on the cases with and without Sommerfeld enhancement in
Figs.~\ref{fig:fSUSY_bounds}-\ref{fig:cross_bounds} are not related to
each other through their values of $\langle\sigma v\rangle_0$.

\subsection{Benchmark models fitting the cosmic ray excesses}\label{bench}

In addition to the reference $\mu^+ \mu^-$ model, we investigate benchmark models recently presented in \cite{Finkbeiner:2010sm}, which produce the correct thermal relic density, fit the cosmic ray (CR) excesses measured by PAMELA and \emph{Fermi}, and are currently allowed by bounds on $S_\mathrm{max}$ from the cosmic microwave background. Whereas our previous reference models demonstrate the effect of Sommerfeld enhancement on the EBL constraints in a broad class of scenarios, these benchmarks allow us to test specific proposed models, and compare our bounds to those from the cosmic microwave background. The parameters characterizing these models are summarized in Table \ref{tab:benchmarks}; see \cite{Finkbeiner:2010sm} for further details \footnote{A web application that computes the Sommerfeld enhancement for this type of models is located at http://astrometry.fas.harvard.edu/mvogelsb/sommerfeld/ (see the last paragraph of the Conclusions in \cite{Finkbeiner:2010sm} for a description).}.

\begin{table*}
\begin{tabular}{|c|c|c|c|c|c|c|}
\hline
Benchmark no. & Annihilation Channel          & $m_\phi$ (MeV) & $m_\chi$ (TeV) & $\alpha_{\rm c}$ & $\delta$ (MeV) & $\frac{S_\mathrm{max} \langle \sigma v \rangle_0}{3 \times 10^{-26} \rm cm^3 \;s^{-1}}$ \\
\hline
1 & 1:1:2 $e^\pm:\mu^\pm:\pi^\pm$ & 900      &  1.68       & 0.04067    & 0.15              & 530         \\
2 & 1:1:2 $e^\pm:\mu^\pm:\pi^\pm$ & 900      &  1.52       & 0.03725    & 1.34             & 360          \\
3 & 1:1:1 $e^\pm:\mu^\pm:\pi^\pm$ & 580      &  1.55       & 0.03523    & 1.49            & 437           \\
4 & 1:1:1 $e^\pm:\mu^\pm:\pi^\pm$ & 580      &  1.20       & 0.03054    & 1.00            & 374          \\
5 & 1:1 $e^\pm:\mu^\pm$           & 350       &  1.33       & 0.02643    & 1.10                 & 339          \\
6 & $e^\pm$ only                  & 200       &  1.00       & 0.01622    & 0.70             & 171          \\
\hline
\end{tabular}
\caption{Particle physics parameters and saturated annihilation cross sections for benchmark points.}
\label{tab:benchmarks}
\end{table*}

The benchmarks feature dark matter masses in the 1-1.7 TeV range, with nearly-degenerate excited states $\delta\sim0.1-1$ MeV above the ground state. Both Sommerfeld enhancement and annihilation to Standard Model final states occur via vector mediators with masses $m_{\phi}$ ranging from $200-900$ MeV. These models were chosen to fit the CR data with \emph{no} contribution from local substructure, so they are not perfectly consistent with the assumptions of this work. However, the Sommerfeld enhancement in these models saturates at relatively high velocities in order to evade constraints from the CMB, and thus we expect the substructure boost to the locally measured CR signals to be only a factor of $\sim 1.5-5$, based on the results of \cite{Kamionkowski:2010mi}. We neglect this effect in the following discussion; if the local substructure boost is substantial then these benchmarks would also significantly overpredict $e^+ e^-$ cosmic rays, and are less interesting for direct comparisons to data.

Since these benchmark models have lower values of $S_{{\rm max}}$ than the off-resonance case
we considered in Fig.~\ref{fig:fSUSY_bounds}, we simply scale-down the upper limits of the latter
to the appropriate value of each benchmark model. We note that although $S(\sigma_{{\rm vel}})$ does
not have the same shape for the benchmark models than for the Yukawa case, the previous
approximation is good enough because most of the signal comes from structures that are already
in the saturated regime. In any case, this approximation actually underestimates the signal
from the benchmark models because $S(\sigma_{{\rm vel}})$ is larger in the intermediate velocity 
dispersions than in the Yukawa case for the same value of $S_{{\rm max}}$.

\begin{figure}
\center{
\includegraphics[height=8.0cm,width=8.0cm]{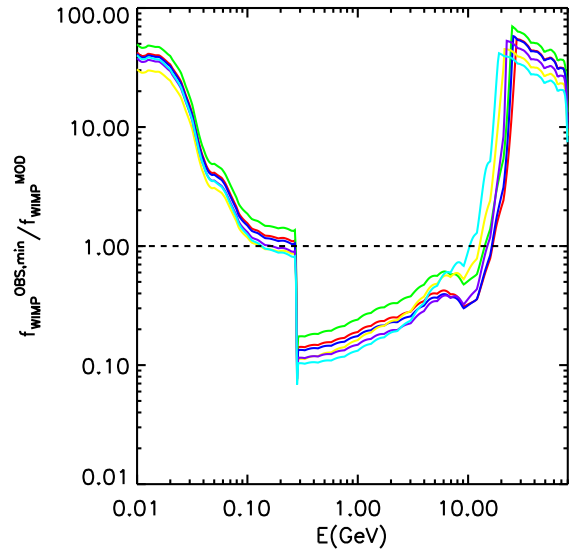}
}
\caption{Ratio of the observed bound on $f_\mathrm{WIMP}$ to the value predicted by the model, for each of the six benchmarks in Table \ref{tab:benchmarks}: 1=red, 2=green, 3=blue, 4=yellow, 5=violet, 6=cyan. Energies where the ratio is less than 1 are ruled out. For the purposes of this figure, we assume minimal contribution from unresolved substructure.}
\label{fig:benchmarks} 
\end{figure}

We find that these benchmarks are in conflict with \emph{Fermi} measurements in the energy range $\sim 0.3-20$ GeV, even in the case of \emph{minimal} contribution from unresolved substructure, if the current best estimates of contributions from blazars and star-forming galaxies to the EBL are subtracted from the data. The conflict is maximal at $E \sim 300$ MeV, where it is a factor of $\sim 7-10$. Fig. \ref{fig:benchmarks} displays the ratio $f_\mathrm{WIMP}^\mathrm{OBS}/f_\mathrm{WIMP}^\mathrm{MOD}$ for these models. With a larger contribution from unresolved substructure, all the benchmarks can be ruled out independent of astrophysical contributions to the EBL.

There are several effects that could ameliorate this conflict, in addition to the small substructure correction mentioned already. The uncertainty on the estimation of $f_\mathrm{WIMP}$, as shown in Fig. \ref{fig:fSUSY_bounds}, can alleviate the tension slightly (by less than a factor of 2); a potentially larger effect is the uncertainty in the subtraction of astrophysical contributions to the EBL. If our best estimate for the contribution of astrophysical sources is too high (by a factor of up to $\sim5$, as discussed previously), then in the case of minimal contribution from unresolved (sub)halos the tension diminishes significantly. In this case however, dark matter annihilation would need to be dominantly responsible for the EBL in the energy range observed by \emph{Fermi}, unless other effects reduce the dark matter signal. 

\begin{figure}
\center{
\includegraphics[height=8.0cm,width=8.0cm]{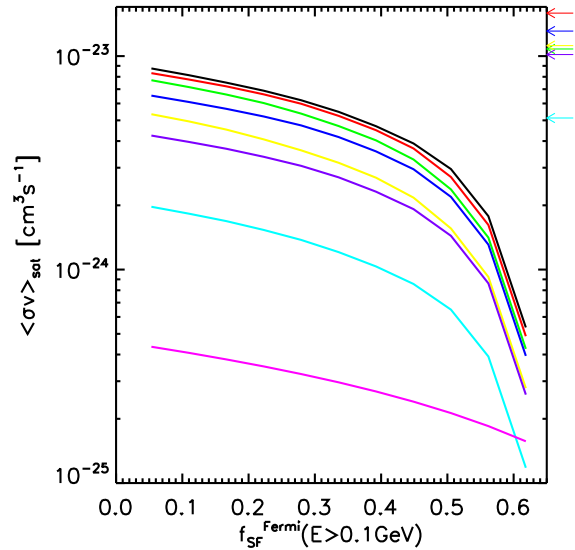}
}
\caption{Constraints to the annihilation cross section at saturation as a function of the 
contribution from star forming galaxies to the observed integrated flux between $0.1$~GeV and $100$~GeV 
as measured by {\it Fermi}. The values above the lines are excluded. We have taken the model 
given in \cite{2009MNRAS.400.2122A} 
to get the spectral shape of this
contribution. We show the six benchmark models of Table 1, setting $\langle\sigma v\rangle_{\rm sat}$ as
a free parameter, with the same colors as in Fig.~\ref{fig:benchmarks}. The small arrows next to
the vertical axis on the right side mark the corresponding values of $\langle\sigma v\rangle_{\rm sat}$ 
for these benchmarks 
as given in Table 1. We also show the results for a model with annihilation into $\mu^+\mu^-$ and a 
mass of $1.5$~TeV (black line),
and the model we used as reference in Figs.~\ref{fig:spectra}-\ref{fig:fSUSY_bounds}: a $185$~GeV 
neutralino annihilating into $b\bar{b}$ (magenta line). As in Fig.~\ref{fig:benchmarks} we have assumed
a minimal contribution from unresolved substructures. Blazars are assumed to contribute by a fixed
amount ($16\%$) in the energy range measured by {\it Fermi} \cite{2010ApJ...720..435A}.}
\label{fig:sigma_sat} 
\end{figure}

Star forming galaxies dominate the astrophysical gamma-ray background model we have used for 
$0.3{\rm ~GeV}\lesssim E_0\lesssim10{\rm ~GeV}$, whereas blazars dominate at higher energies. The contribution of the
former to the total observed signal is particularly important to constrain the role of dark matter 
annihilation. To illustrate this, we translate the constraints on 
$f_{\rm WIMP}$ given in Fig.~\ref{fig:fSUSY_bounds} to constraints on the value of the annihilation 
cross section at saturation by taking the value of 
$\langle\sigma v\rangle_{{\rm sat}}=S_{\rm max}\langle\sigma v\rangle_{0}$ as a free parameter limited by 
the {\it Fermi} measurements and the astrophysical background. Fig.~\ref{fig:sigma_sat} shows these 
constraints as a function of $f_{\rm SF}^{\rm Fermi}(E>0.1 {\rm GeV})$, the contribution of star forming 
galaxies to the observed integrated flux between $0.1$~GeV and $100$~GeV. We show the constraints only 
for the case of
minimal contribution of unresolved subhalos. The six benchmark models appear with the same colors as in 
Fig.~\ref{fig:benchmarks}. As a reference, the values of 
$\langle\sigma v\rangle_{{\rm sat}}$ that fit the cosmic ray excesses for these models are marked with arrows
next to the vertical axis on the right side. The other two
models we have used throughout the paper are also included in the figure: $m_{\chi}=185$~GeV 
annihilating into $b\bar{b}$ (magenta line) and $m_{\chi}=1.5$~TeV annihilating into $\mu^+\mu^-$
(black line). For the benchmark models and for the $\mu^+\mu^-$ model, the constraint on 
$\langle\sigma v\rangle_{{\rm sat}}$ decreases rapidly with $f_{\rm SF}^{\rm Fermi}(E>0.1 {\rm GeV})$,
because these models are constrained at $E_0\sim0.3$~GeV where the contribution from star forming galaxies
peaks. Even assuming only a $5\%$ contribution of star 
forming galaxies (recall that we have used $53\%$ as a fiducial value), the constraints on 
$\langle\sigma v\rangle_{{\rm sat}}$ still exclude the values needed by these models to fit the cosmic ray excesses.
The $b\bar{b}$ model is more independent of the star forming contribution since this model is constrained at
$E_{0}\sim10$~GeV, where blazars dominate.

We would like to mention that the model we have used to include the contribution from star forming galaxies assumes 
that the Milky-Way gamma-ray specific intensity has a power law energy spectrum with an exponent of $-2.7$ for $E\geq0.6$~GeV 
\cite{2009MNRAS.400.2122A}, which seems to be too steep at high energies according to the recent analysis of the 
{\it Fermi-LAT} collaboration that points to an exponent close to $-2.5$ for $E\geq10$~GeV (see Table I and 
Fig.~3 of \cite{2010PhRvL.104j1101A}). Assuming a shallower spectrum for the Milky-Way gamma-ray specific intensity
would result in a contribution of star forming galaxies to the EBL with a shallower spectrum as well, making it more relevant 
at higher energies that it is in the model we have used in this work, and slightly strengthening the derived constraints in 
Figs.~\ref{fig:EBL_Somm},\ref{fig:fSUSY_bounds} and \ref{fig:benchmarks} at high energies. Nevertheless, the most relevant 
uncertainty is the overall contribution of star forming galaxies discussed in the previous paragraph and whose
effects are shown in Fig.~\ref{fig:sigma_sat}.

We have assumed a low-mass cutoff of $10^{-6}$ $M_\odot$: kinetic decoupling can occur quite late in models of this type \cite{Feng:2010zp}, leading to a higher cutoff of up to $0.1-1$ $M_\odot$ \cite{Bringmann:2009vf}. However, we estimate that a change of five orders of magnitude in the low-mass cutoff will affect the final result by a factor of only $2-6$, and the high end of this range will only be attained for non-minimal contributions from unresolved (sub)halos (i.e., scenarios that are presently in conflict with even the unsubstracted data, for these benchmark models). If the slope we have assumed for the central density profile of the halos is steeper than reality, this could also affect our limits by a factor of a few: the NFW profile we have chosen lies between the Moore and Burkert profiles considered by \cite{2009JCAP...07..020P}, and the difference between those profiles modifies the gamma-ray signal by roughly an order of magnitude. The presence of significant dark matter self-interactions and nearly-degenerate excited states in models of this type can lead to disruption of low-mass halos and the depletion of central density cusps (see e.g. \cite{Loeb:2010gj,Bell:2010fk} and references therein); while these effects could potentially reduce the tension with the EBL data, their inclusion is beyond the scope of our current analysis.

We therefore see that in a CDM scenario, in the context of current structure formation models, the EBL can robustly act as a more sensitive probe of Sommerfeld-enhanced dark matter annihilation scenarios than the cosmic microwave background. Removing tension with the EBL for the benchmark models we have tested seems to demand minimal contributions to the signal from unresolved substructure, and \emph{in addition} either dark matter structure formation must be modified from a pure collisionless CDM scenario, or the contribution to the EBL from blazars and galaxies must be at the low end of current estimates.

\section{Summary and Conclusions}\label{concl}

A positive detection of a non-gravitational signature of dark matter
would be a breakthrough in our understanding of this still mysterious
form of matter.  Current experiments on Earth looking for signals of
interactions between dark and ordinary matter intensify the
efforts to reach the necessary sensitivities to either largely
constrain the parameter space of minimal SUSY theories that predict
the favorite dark matter candidate, the neutralino, or to find a
definite signal \cite{2010arXiv1009.3934F}.

The existence of dark matter could also be confirmed through the
detection of ordinary matter produced during the annihilation of
WIMPs in regions of high dark matter density. This annihilation
is expected to produce a population of gamma-ray photons that would
make dark matter halos visible in the gamma-ray sky. The cumulative
effect of these gamma-rays produced outside our galactic halo creates
a cosmic background that adds up to the one produced by other sources
such as blazars and star forming galaxies.

This hypothetical background radiation is also populated at lower
energies by a fraction of the original CMB photons that on their
journey towards us are scattered by energetic electrons and positrons
produced during the annihilation of WIMPs. They gain energy in
the process and reach us as X-ray and gamma-ray photons.

In this work, we have used the state-of-the-art Millennium II
simulation \cite{2009MNRAS.398.1150B} that follows the formation and
evolution of structure formation in a $\Lambda$CDM cosmology, to
produce simulated sky maps of this conjectured cosmic background. Our
method includes the signal coming from all halos and subhalos
resolved in the simulation as well as a careful extrapolation to
account for the contribution of unresolved structures that are
expected to exist all the way down to masses of about $\sim 1$ Earth mass, 
that correspond to the damping mass limit of one of the most studied 
type of WIMPs: $\sim100$~GeV neutralino.

This paper extends the analysis of \cite{2010MNRAS.tmp..453Z} by
including: i) the X-ray and soft gamma-ray contribution to the
background radiation by CMB photons that gain energy through Inverse
Compton scattering of the electrons and positrons produced during
annihilation \footnote{Additional backgrounds such as starlight,
  infrared and ultraviolet contribute as well, but we have neglected
  them since the CMB is by far the largest of these background
  populations.}; ii) a detailed treatment of a Sommerfeld mechanism
that enhances the annihilation cross section, leading to a
significantly larger annihilation rate from dark matter structures
with low velocity dispersions. The Sommerfeld enhancement has been
invoked to explain the anomalous excess of cosmic ray positrons above
10~GeV reported by the PAMELA satellite
(e.g. \cite{2009PhRvD..79a5014A}). We present results using this
enhancement for two sets of parameters chosen to represent typical
cases: i) an off-resonance case where the boost to the annihilation
cross section scales as $1/\sigma_{\rm vel}$, and ii) a near-resonance
case where the boost goes as $1/\sigma_{\rm vel}^2$.

We have found that observational upper limits on the unknown
contributions to the X-ray and gamma-ray background radiation put
significant constraints on the contribution from dark matter
annihilation (see Fig.~\ref{fig:EBL_Somm} for a comparison with a
particular model). These upper limits are especially stringent in the
gamma-ray regime due to recent measurements reported by the {\it Fermi-LAT}
experiment, together with well-founded expectations for the
contributions of blazars and star-forming galaxies
\cite{2010JCAP...04..014A,2010ApJ...720..435A,2009MNRAS.400.2122A}.

We introduced a model-independent way to give constraints on the
intrinsic properties of WIMPs by ``factoring-out'' the
astrophysical part of the signal, namely, the one that depends on the
density field of dark matter, which is accurately given by the N-body
simulation we have used. The constraints we obtain for the remaining
``particle physics'' factor ($f_{\rm WIMP}$, see Eq.~\ref{emiss}), appear on
Fig.~\ref{fig:fSUSY_bounds}. This figure can be used as a template to
test whether or not a given particle physics model violates the
observational constraints. Although for the case with Sommerfeld
enhancement we only presented two particular cases,
Fig.~\ref{fig:fSUSY_bounds} can still be easily used to scale the
constraints up or down for other realizations of these types of
models.

By selecting a particle physics model and computing the photon yield
${\rm d}N/{\rm d}E$ (composed by in situ and up-scattered CMB photons, see
sections \ref{in_situ}, \ref{up_scat} and Fig.~\ref{fig:spectra}), it
is possible to give direct constraints for the annihilation cross
section as a function of WIMP mass. We show an example of this
in Fig.~\ref{fig:cross_bounds}, where a model annihilating into
$\mu^{+}\mu^{-}$ final states was chosen. For this particular model,
the constraints we obtain disfavor the scenario where the positron
excess measured by the PAMELA satellite is explained by dark matter
annihilation alone (of course, there could still be some subdominant DM contribution to the signal). Furthermore, we have presented constraints on specific ``benchmark'' Sommerfeld-enhanced models selected to fit the cosmic ray spectra measured by PAMELA and \emph{Fermi} without any contribution from local substructure, while obtaining the correct relic density and respecting bounds from the cosmic microwave background.
We find that these models are in conflict with our constraints, even in the case of minimal contributions from unresolved substructure. This tension could diminish significantly if the contribution to the cosmic gamma-ray background from blazars and star forming galaxies is quite low (current uncertainties are still large, particularly in the latter, and put a minimum value of $17\%$ of the observed signal at $E\sim0.3$~GeV which is a factor of 
5 lower
than the estimate we have used here, see the last paragraph of section \ref{obs}). Another interesting
possibility to reconcile these models lies in taking into account the role of 
self-interactions between dark matter particles, inherent in the models, in the formation
and evolution of dark matter structures (this can lead for example to the formation of central density cores in 
low-mass halos \cite{Loeb:2010gj}). 

The main sources of uncertainty in our modelling from the
astrophysical part of the signal are, in order of importance: i) the
contribution of unresolved substructures, which is uncertain by
roughly two orders of magnitude; ii) the concentration of dark matter
in the inner part of halos. In this work we have used a NFW density
profile. If an Einasto profile is used instead, which is currently
favored by high resolution simulations of single halos, the
annihilation rate for each halo is increased by $50\%$
\cite{2010MNRAS.tmp..453Z}; iii) the value of the minimum mass of
bound halos made of WIMPs; iv) the approximations used for electron and
positron losses, and photon absorption (see Appendices A and B).

It is worth mentioning that the fine-grained structure of dark matter halos is predicted to be a superposition 
of streams with very small internal velocity dispersions. If
the annihilation cross section is independent of the velocity dispersion, then the contribution of these fundamental
streams and their associated caustics to the annihilation rate is essentially negligible \cite{2010arXiv1002.3162V}. 
However, this could change dramatically in Sommerfeld-enhanced models due to the large boosts expected in the streams.
We explore this in appendix \ref{streams_sec} and find that despite the more prominent role of streams in these type
of models, their contribution is still significantly smaller than that of subhalos, due to the saturation of the
enhancement at low velocities, and can be safely neglected.

In spite of these uncertainties, and thanks to increasingly better
measurements of the cosmic X-ray and gamma-ray background radiation,
and to our better understanding of the contribution to it by AGNs,
blazars and star-forming galaxies, an analysis like ours produces
competitive constraints compared to those obtained in other indirect
searches, such as those based on dwarf galaxies
\cite{2010arXiv1007.4199E}.  Our work can also be viewed as
complementing that of other works
\cite{2009JCAP...07..020P,2010PhRvD..81d3505B,2010JCAP...07..008H},
that have presented a similar analysis using analytical approaches to model
the astrophysical part of the signal instead of high resolution N-body
simulation, as we have done here.

\section*{Acknowledgments}

We would like to thank Neal Weiner, Simon D. M. White, Niayesh Afshordi and 
Mattia Fornasa for interesting comments and suggestions. JZ acknowledges 
financial support by the Joint Postdoctoral Program in
Astrophysical Cosmology of the Max Planck Institute for Astrophysics
and the Shanghai Astronomical Observatory, and  
from a CITA National Fellowship. This work was supported in
part by NSF grant AST-0907890 and NASA grants NNX08AL43G and
NNA09DB30A (for A.L.). TS gratefully acknowledges support from the Institute for Advanced Study. 
The research of TS is supported by the DOE under grant
DE-FG02-90ER40542 and the NSF under grant AST-0807444.

\appendix

\section{Energy losses for electrons and positrons}

The processes briefly summarized here are described in detail in
\cite{2002cra..book.....S}. Since the CMB energy density scales with
redshift as $(1+z)^4$, the energy loss term due to IC scattering with
the CMB photons is given by:
\begin{equation}\label{loss_IC}
b_{\rm e}(E_{\rm e},z)_{\rm ic}\approx2.5\times10^{-17}(1+z)^4\left(\frac{E_{\rm
    e}}{{\rm GeV}}\right )^2\ {\rm GeV}/{\rm s}
\end{equation}

The energy loss due to synchrotron radiation in the ambient magnetic field
$B$, which has a spatial and temporal functional dependence, is given by:
\begin{equation}\label{loss_S}
b_{\rm e}(E_{\rm e},z)_{\rm syn}\approx0.254\times10^{-17}\left(\frac{B}{{\rm 1\mu
    G}}\right )^2 \left(\frac{E_{\rm e}}{{\rm GeV}}\right )^2\ {\rm GeV}/{\rm s}
\end{equation}
The magnitude of the magnetic field has large spatial variations,
going from $\sim10\mu$G in the cores of galaxy clusters
\cite{2002ARA&A..40..319C} to $\sim0.1\mu$G in the intergalactic
medium in clusters \cite{2007ApJ...659..267K}. Thus, synchrotron
losses are expected to be comparable to IC losses only in the regions
with the strongest magnetic fields, such as the centers of galaxy
clusters. Although dark matter annihilation is copious in high density
regions such as these, the contribution from subhalos and low-mass
halos is in average more significant than the one from the center of
massive halos associated to galaxy clusters. Furthermore, the
strength of the magnetic field is not expected to increase as rapidly
with redshift as the CMB energy density. This makes the synchrotron
losses less significant than the IC losses at high redshifts.

The electrons and positrons produced in the annihilation process also
lose energy due to ionisation of neutral atoms and Coulomb scattering
with free electrons present in the ambient field. The energy loss rate
of both processes is essentially independent of energy and is given
by:
\begin{eqnarray}\label{loss_ion_coul}
b_{\rm e}(E_{\rm e},z)_{\rm
  ion}\approx18.4\times10^{-17}\left(\frac{n_{\rm H}}{{\rm
    cm}^{-3}}\right)\ {\rm GeV}/{\rm s}\\ b_{\rm e}(E_{\rm e},z)_{\rm
  coul}\approx55.4\times10^{-17}\left(\frac{n_{\rm e}}{{\rm
    cm}^{-3}}\right)\ {\rm GeV}/{\rm s}
\end{eqnarray}
where $n_{\rm H}$ and $n_{\rm e}$ are the local number densities of neutral
hydrogen and free electrons, respectively. Bremsstrahlung radiation is
another source of energy loss that also depends on the local density
of the ambient ionized and neutral material. In the weak-shielding
limit the energy loss rate due to Bremsstrahlung is given by:
\begin{equation}\label{loss_B}
b_{\rm e}(E_{\rm e},z)_{\rm brem}\approx15.1\times10^{-17}\left(\frac{E_{\rm e}}{{\rm
    GeV}}\right )\left(\frac{n_{\rm e}}{{\rm cm}^{-3}}\right)\ {\rm GeV}/{\rm s}
\end{equation}

At high electron energies, the latter three processes are subdominant
relative to the IC losses due to the energy squared dependence in
Eq.~(\ref{loss_IC}). At low energies they become more significant but
are nevertheless suppressed by the average low density of the ambient
medium. This can be seen by noting that the minimum energies we are
interested in are those corresponding to the soft X-rays
($E_{\rm IC}\gtrsim10^{-7}$GeV) coming from the IC scatter of
CMB photons. An up-scattered CMB photon will have an average energy of:
$E_{\rm IC}\approx4/3(E_{\rm e}/m_{\rm e})^2E$ \cite{1996AN....317..156L}, where
$E_{\rm e}$ and $m_{\rm e}$ are the energy and mass of the scattering electron,
and $E$ is the energy of the photon before the event, thus for $E_{\rm
  IC}\sim10^{-7}$~GeV the electron energies of relevance are of the
order of $0.25$~GeV at $z=0$. For these energies, ionisation, coulomb
and bremsstrahlung losses dominate over IC losses only if the ambient
density of electrons and neutral hydrogen is
$\gtrsim10^{-2}$~cm$^{-3}$. In the local ISM $n_{\rm e}\sim0.1$~cm$^{-3}$
\cite{2008ApJ...683..207R}. In galaxy clusters the average gas
density is $\sim10^{-3}$~cm$^{-3}$ \cite{2006A&A...455...21C} and in
dwarf spheroidals like Draco it is $\sim10^{-6}$~cm$^{-3}$
\cite{2007PhRvD..75b3513C}.  Since the largest contribution to the
production of electrons and positrons comes from the accumulated
effect of annihilation in low-mass halos and subhalos, which have
clearly low ambient densities of ordinary matter, we can safely
neglect the impact of these three processes of energy loss. They could
be of relevance in the center of massive halos at low redshift, but
they are negligible for the overall full-sky signal.

\section{Photon absorption}

For energies $\gtrsim10$~GeV measured at $z=0$, the dominant
mechanism of photon absorption is that due to the interaction between
the gamma-ray photons produced in the annihilation process and the
lower energy starlight photons produced in galaxies (i.e., pair
production with the ambient photon field). As mentioned in Section
\ref{formalism}, this absorption is parameterised as an exponential
term with an effective optical depth $\tau(E_{0},z)$. We adopt the
most recent treatment of \cite{2009MNRAS.399.1694G} to calculate the
values of the optical depth as a function of energy and redshift. For
this purpose, we take their fiducial 1.2 model and make a bilinear
interpolation following their Fig. 11.

For lower observed energies down to $\sim10^{-6}$~GeV, the
Universe is basically transparent to photons produced at any given
redshift between $z=0$ and $z=10$ (e.g. Fig. 3
  of \cite{2010JCAP...07..008H}).  In this paper we are considering a
range of energies that extends slightly towards lower energies
($10^{-7}$~GeV). In this regime, photoionization and Compton
scattering are important mechanisms of energy loss. As can be seen
from Fig. 3 of \cite{2010JCAP...07..008H}, $\tau\sim 1$ at these
energies for photon sources located at $z\sim7$.  This means that
these processes would suppress an important fraction of photons coming
from dark matter structures at $z\gtrsim7$. However, most of the
emission from annihilation comes from sources at $z\lesssim3$
($\sim60\%$ of the total emission at these energies), which is a
region essentially transparent. Thus, we are ignoring these mechanisms
noting that we could be overestimating the predicted signal at the
percent level, which is clearly a minor effect for the purposes of
this work.

\section{The astrophysical factor, luminosity from halos and
  subhalos}\label{astro_factor}

\subsection{Resolved structures}

The total annihilation luminosity (including in situ and up-scattered
photons) coming from a halo (or subhalo) of volume V is given by:
\begin{eqnarray}\label{lum1}
L_{\rm h}&=&\int_{\rm V}\mathcal{E}({\vec x})~{\rm d}V =\frac{E}{2}f_{\rm
  WIMP}\int_{\rm V}\rho_{\chi}({\vec x})^2 S(\sigma_{\rm vel}({\vec x}))~{\rm
  d}V \nonumber \\ &=&\frac{E}{2}f_{\rm WIMP}L_{\rm h}'.
\end{eqnarray}
We assume that halos (or subhalos) have a NFW density profile\footnote{An Einasto
profile seems to be favored over a NFW by the most recent
high-resolution N-body simulations
(e.g. \cite{2010MNRAS.402...21N}), 
using the former instead of the latter increases the net
annihilation rate in a halo by only $50\%$ \cite{2010MNRAS.tmp..453Z}. This effect is small
compared to other uncertainties in our analysis.}
\cite{1997ApJ...490..493N}, and an average boost factor
$S(\bar{\sigma}_{\rm vel})$ given by the mean velocity dispersion of
its particles. Thus:
\begin{equation}\label{scaling_law}
L_{\rm h}'=S(\bar{\sigma}_{\rm vel})\int\rho^2_{\rm NFW}(r)\,{\rm
  d}V=S(\bar{\sigma}_{\rm vel})\frac{1.23\,V_{\rm max}^4}{G^2r_{\rm
    max}},
\end{equation}
where the last scaling relation was found by
\cite{2008Natur.456...73S} with $r_{\rm max}$ being the radius where
the rotation curve reaches its maximum $V_{\rm max}$. It is important to 
consider the impact of numerical resolution on
the values of $r_{\rm max}$ and $V_{\rm max}$. 
The values of $r_{\rm max}$ are increasingly overestimated for smaller
structures whereas the opposite is true for $V_{\rm max}$ 
\cite{2008MNRAS.391.1685S,2010MNRAS.tmp..453Z}.  We have
hence corrected these quantities following the prescription of \cite{2010MNRAS.tmp..453Z}.

\subsection{Unresolved structures}

The previous description is used for all structures that are resolved
by the MS-II simulation. However, we want to obtain predictions down
to the minimum mass for bound WIMP halos. For neutralinos this is $\lesssim
1$M$_{\oplus}$, which clearly lies many orders of magnitude below
current simulations.  For $\sim100$~GeV neutralinos, the damping mass lies
in the range $10^{-8}-10^{-4}$ M$_{\odot}$, whereas for
$\sim1$~TeV neutralinos, the range is $10^{-11}-10^{-7}$ M$_{\odot}$. For
simplicity, we assume that these reference values for neutralinos are generically valid for other
WIMPs and choose a fiducial value of $10^{-6}h^{-1}$M$_{\odot}$ for
all the cases we analyze in this paper, noting that the
precise value of this mass is a source of uncertainty in our results.

To incorporate these unresolved structures into our maps, we follow an
analogous procedure to the one developed in
\cite{2010MNRAS.tmp..453Z}, that we briefly describe in the following,
dividing it into unresolved main halos and unresolved subhalos.

\subsubsection{Unresolved halos} 

The total annihilation luminosity coming from main halos in a given
mass range can be computed using the function:
\begin{equation}\label{F_M}
F_{\rm h}(M_{\rm h})=\frac{\sum L_{\rm h}'}{\bar{M}_{\rm h}\Delta \log M_{\rm h}},
\end{equation}
where the sum is over all the luminosities $L_{\rm h}'$ of halos (given by
Eq.~\ref{scaling_law}) with masses in the range: $\log
M_{\rm h}\pm \Delta \log M_{\rm h}/2$, and $\bar{M}_{\rm h}$ is the
mean halo mass in each bin. In the absence of Sommerfeld enhancement
the function $F_{\rm h}(M_{\rm h})$, henceforth called: $F^{\rm
  NSE}_{\rm h}(M_{\rm h})$, is a power law in the intermediate to low
mass regime (see Fig.~4 of \cite{2010MNRAS.tmp..453Z}).  
Once the Sommerfeld boost is applied to each main halo, the power law
behavior of $F_{\rm h}(M_{\rm h})$ is modified by the function
$S(\bar{\sigma}_{\rm vel})$. 

The minimum halo mass we can rely on to
compute $F_{\rm h}(M_{\rm h})$ is $M_{\rm
  lim}=6.89\times10^8\hMsol$ (100 simulation particles). Below
this mass we need to extrapolate $F_{\rm h}(M_{\rm h})$ using the
information we have on $S(\bar{\sigma}_{\rm vel})$, and on the
extrapolation made for $F^{\rm NSE}_{\rm h}(M_{\rm h})$.  The value of
$M_{\rm lim}$ translates into a limiting value of $\sigma_{\rm vel}$
that we obtain directly from the simulation data: $\sigma_{\rm
  vel,lim}(z=0)\sim3.4\times10^{-5}$ \footnote{At z=0, $\sigma_{\rm vel}\propto
M_{\rm v}^{0.34}$}. Therefore, we
obtain a fit to the power law behavior of $F_{\rm h}(M_{\rm h})$ in
the last resolved mass range of the MS-II and extrapolate this
function down to the damping mass limit taking into account the saturation of
$S(\bar{\sigma}_{\rm vel})$.

The ratio of annihilation emission coming from all halos contained in
a cosmic volume $V_{\rm B}$ with masses larger than $M_{\rm min}$ to the
emission produced by a smooth homogeneous distribution of dark matter,
with average density $\bar{\rho}_{\rm B}$, filling this volume is
approximately given by:
\begin{equation}\label{flux_mmin_sim}
  f(M_{\rm h}>M_{\rm min})\sim\frac{1}{\bar{\rho}_{\rm B}^2V_{\rm B}}\int_{M_{\rm
      min}}^{\infty}\frac{F_{\rm h}(M_{\rm h})}{\ln 10}\,{\rm d}
  M_{\rm h} .
\end{equation}

For a given redshift, the ratio of the values of $f(M_{\rm h}>M_{\rm
  min})$ with and without enhancement below the saturation mass is
roughly given by $S_{\rm max}$.

Using Eq.~(\ref{flux_mmin_sim}), we estimate the contribution from
the unresolved main halos down to the damping mass limit by
assuming that the radiation from the missing halos in the mass range
$10^{-6}\hMsol$ to $\sim6.89\times10^8\hMsol$ is distributed on the
sky in the same way as the one from the smallest masses we can resolve
in the simulation, which we adopt as the mass range between
$1.4\times10^{8}\hMsol$ and $\sim6.89\times10^8\hMsol$ (halos with 20
to 100 particles). This assumption is justified because the clustering
bias seems to asymptotically approach a constant value for low halo
masses \cite{2009MNRAS.398.1150B}.

Using the extrapolated behavior of $F_{\rm h}(M_{\rm h})$ in
Eq.~(\ref{flux_mmin_sim}) we compute the boost factor $b_{\rm h}$ by
which each halo in the mass range $1.4-6.89\times10^8\hMsol$ needs to
be multiplied such that the luminosity of the unresolved main halos
is accounted for as well:
\begin{eqnarray}\label{boost_h}
b_{\rm
  h}^{({\rm NSE},i,ii)}&=&\frac{f(10^{-6}\hMsol,6.89\times10^8\hMsol)_{\rm a}}{f(1.4\times10^{8}\hMsol,6.89\times10^8\hMsol)_{\rm
    sim}}\nonumber\\ &\sim& (60,90,2.4\times10^3)
\end{eqnarray}
Note that $b_{\rm h}$ is effectively the ratio of $f(M_{\rm h}>M_{\rm
  min})$ computed analytically between the cutoff mass limit and the
100 particle limit, and computed in the simulation for the lowest
resolved mass range. The superscripts $(NSE,i,ii)$ are for the cases 
without Sommerfeld enhancement, off-resonance ($S_{\rm max}=2000$) and 
near-resonance ($S_{\rm max}=10^6$), respectively. 
The value of $b_{\rm h}$ is
nearly independent of redshift up to $z=2.1$. For higher
redshifts, the power law fit to Eq.~(\ref{F_M}) is unreliable for the extrapolation
because the population of halos over the resolved mass range becomes too small.

\subsubsection{Unresolved subhalos }\label{un_subs}

Cold dark matter halos contain numerous substructures that contribute
significantly to their total annihilation luminosity. For massive
halos, this contribution largely exceeds that of the smooth main
halo. For a MW-like halo
the total luminosity from all its subhalos down to the damping mass
is between $2$ and $2000$ times larger than its own
smooth component \cite{2010MNRAS.tmp..453Z}.

The Sommerfeld mechanism increases the contribution of substructures
even further due to their low velocity dispersion relative to that of
their host. We now calculate the contribution from unresolved
subhalos following an analogous procedure to the one described in
\cite{2010MNRAS.tmp..453Z}. It follows a methodology similar to that
of the previous subsection and rests on the analysis of the following
quantity:
\begin{equation}\label{F_subs}
F_{\rm sub}\left(\frac{M_{\rm sub}}{M_{\rm
    h}}\right)=\left(\frac{M_{\rm h}}{L_{\rm h}'}\right)\frac{\sum
  L_{\rm sub}'}{\bar{M}_{\rm sub}\Delta \log M_{\rm sub}},
\end{equation}
where $M_{\rm sub}$ and $L'_{\rm sub}$ (given by Eq.~(\ref{scaling_law}) are the mass and luminosity of
a given subhalo. The total
luminosity of all subhalos relative to that of their host is given
by:
\begin{equation}\label{subs_extrapol}
f_{\rm sub}(M_{{\rm sub}}^{{\rm max}},M_{\rm h})\sim\frac{1}{L_{\rm
    h}'}\int_{10^{-6}}^{M_{{\rm sub}}^{{\rm max}}}\left(\frac{L_{\rm
    h}'}{M_{\rm h}}\right) \frac{F_{\rm sub}\left(\frac{M_{\rm
      sub}}{M_{\rm h}}\right)}{\ln 10}{\rm d}M_{\rm sub} ,
\end{equation}
where $M_{{\rm sub}}^{{\rm max}}$ is the mass of the most massive
subhalo within the host.

To simplify the analysis of $F_{\rm sub}$ in the case of Sommerfeld
enhancement, we approximate $F_{\rm sub}$ by:
\begin{equation}\label{f_sub_2}
 F_{\rm sub}\!\sim\!\left(\frac{M_{\rm h}}{\bar{S}(M_{\rm h})L_{\rm
     h}'^{\rm NSE}}\right)\frac{\bar{S}(M_{\rm sub})\sum L_{\rm
     sub}'^{\rm NSE}}{\bar{M}_{\rm sub}\Delta \log M_{\rm sub}}\!=\!
 \frac{\bar{S}(M_{\rm sub})}{\bar{S}(M_{\rm h})}F_{\rm sub}^{\rm NSE}\\ 
\end{equation}
where $\bar{S}(M_{\rm h})$ and $\bar{S}(M_{\rm sub})$ are average
enhancements for $M_{\rm h}$ and for the subhalo mass
range: $\log M_{\rm sub}\pm \Delta \log M_{\rm sub}/2$,
respectively. These averages are given by the combination of
$\sigma_{\rm vel}(M)$ and $S(\sigma_{\rm vel})$, and we obtain them
directly by fitting the simulation data.

The average boost as a function of halo mass and redshift is well
described by:
\begin{equation}\label{S_M}
(\bar{S}(M_{\rm h},z))_{(i,ii)}=(S_{\rm h,0}G_{\rm h}(z)M_{\rm
    h}^{\alpha_{\rm h}^{\rm SE}})_{(i,ii)},
\end{equation}
where $S_{\rm h,0}$ is a normalization factor and all redshift
dependence has been put into $G_{\rm h}(z)$ (recall that this
dependence comes from the $\sigma_{\rm vel}(M_{\rm h})$ relation).

For case i), there are two characteristic masses that mark the
transitions below which $S\rightarrow S_{\rm max}$ and above which
$S\rightarrow 1$. The characteristic masses are redshift dependent and
can be obtained by matching the three regimes. For case ii), there are
three such characteristic masses marking the transition between
saturation, $S\sim1/\sigma_{\rm vel}^2$, $S\sim1/\sigma_{\rm vel}$ and
$S\rightarrow 1$. For both cases we find that the fitting functions
are a very good approximation up to $z=2.1$.

For subhalos we apply a similar procedure using:
\begin{equation}\label{S_sub}
(\bar{S}(M_{\rm sub},z;M_{\rm h}))_{(i,ii)}=(S_{\rm sub,0}G_{\rm
    sub}(z)M_{\rm sub}^{\alpha_{\rm sub}^{\rm SE}})_{(i,ii)},
\end{equation}
which has the same functional form as Eq.~(\ref{S_M}) but with an
implicit dependence on the mass of the host which takes care of the
fact that subhalos can only have Sommerfeld boosts that are larger
than those of their hosts.  Thus, if for example a host is saturated,
all its subhalos are saturated as well and we have $\bar{S}_{\rm
  sub}=\bar{S}_{\rm h}$. For this particular case it is easy to see
that the contribution of substructures to the luminosity of the halo
is the same as in the case without enhancement: $f_{\rm sub}=f_{\rm
  sub}^{\rm NSE}$. To obtain the parameters in Eq.~(\ref{S_sub}) we
only analyze main halos with more than 500 subhalos.

In both cases the different regimes are divided by transition subhalo
masses analogous to the ones for hosts. The fitting procedure is less
reliable than in the case of halos. In particular, the scatter on the
slope of the power law for a given redshift, measured with quartiles,
is of the order of $10\%$, and the median can change up to the same
amount between $z=0$ and $z=2.1$. The scatter of the normalization at
a given redshift is of the order of a factor of 2, and for different
redshifts, its median can change up to a factor of~4.

As for halos, we consider the subhalo population
to be complete down to $M_{\rm lim}=6.89\times10^8\hMsol$. Below this mass, we
use Eqs.~(\ref{subs_extrapol}-\ref{S_sub}) to add the
contribution of unresolved subhalos to each of the resolved halos. 
The value of $M_{\rm sub}^{\rm max}$ in Eq.~(\ref{subs_extrapol}) is
given by $M_{\rm sub}^{\rm max}=M_{\rm lim}$ if the halo has subhalos and
$M_{\rm sub}^{\rm max}=f_{\rm max}M_{\rm h}$, with $f_{\rm max}=0.05$
otherwise. In the former case we distribute the missing luminosity
among all resolved subhalos, in the latter we simply add it to the
host. The precise value of $f_{\rm max}$ has little impact on the
results. 

Considering resolved an unresolved subhalos we find that a halo of $10^{12}\hMsol$ has $f_{\rm
  sub}\in(11,1.04\times10^{4})$ for case i), and $f_{\rm
  sub}\in(4.13\times10^3,6.86\times10^{6})$ for case ii), that is $\sim6$,
$\sim3000$ times more than in the case without Sommerfeld enhancement,
respectively.

Finally, we need to add the subhalo contribution to all main halos
with masses below $M_{\rm lim}$. To do so, we compute an overall boost
factor $b_{\rm sub}$ to the luminosity of all main halos between the
damping scale limit and $M_{\rm lim}$:
\begin{equation}\label{boost_subs}
  b_{\rm sub}=\frac{\mathfrak{f}_{\rm
      boost}(10^{-6}\hMsol,6.89\times10^8\hMsol)}{\mathfrak{f}_{\rm
      no-boost}(10^{-6}\hMsol,6.89\times10^8\hMsol)}
\end{equation}
where $\mathfrak{f}_{\rm no-boost}$ is given by:
\begin{align}\label{boost_subs_1}
&\mathfrak{f}_{\rm
    no-boost}(10^{-6}\hMsol,6.89\times10^8\hMsol)\approx\nonumber\\ &\int_{10^{-6}}^{6.89\times10^8}\frac{F_{\rm
      h}(M_{\rm h})}{\ln 10}\,{\rm d}M_{\rm h}
\end{align}
and $\mathfrak{f}_{\rm boost}$ can be written as:
\begin{align}\label{boost_subs_2}
&\mathfrak{f}_{\rm
    boost}(10^{-6}\hMsol,6.89\times10^8\hMsol)\approx\nonumber\\ &\int_{10^{-6}}^{6.89\times10^8}[1+f_{\rm
      sub}(f_{\rm max} M_{\rm h},M_{\rm h})]\frac{F_{\rm h}(M_{\rm
      h})}{\ln 10}\,{\rm d}M_{\rm h} .
\end{align}

For the cases with Sommerfeld enhancement, Eq.~(\ref{boost_subs}) can be simplified 
by noting the following. The
integral in Eq.~(\ref{boost_subs_1}) is dominated by the mass range
where the Sommerfeld enhancement is already saturated, this is because
$F_{\rm h}$ is always a power law, monotonically increasing with mass,
and the saturation mass ($M_{\rm h,sat}$) is relatively close to
$M_{\rm lim}$ and much larger than the damping mass. For instance in
case i), $M_{\rm h,sat}\sim7\times10^8\hMsol(1+z)^{-1}$, thus even at
high redshifts, the contribution of the unsaturated part is always
negligible. In case ii), $M_{\rm
  h,sat}\sim1.5\times10^5\hMsol(1+z)^{-0.8}$, which means that the
aforementioned contribution is larger than in case i) but we still
find it to contribute minimally to the integral, less than $10\%$.
Therefore, we can approximate Eq.~(\ref{boost_subs_1}) by:
$\mathfrak{f}_{\rm no-boost}\approx S_{\rm max}\mathfrak{f}_{\rm
  no-boost}^{\rm NSE}$, where $\mathfrak{f}_{\rm no-boost}^{\rm NSE}$
is the value of Eq.~(\ref{boost_subs_1}) in the case of no Sommerfeld
enhancement. For case i), a similar approximation can be used to
simplify Eq.~(\ref{boost_subs_2}). It can be shown that $f_{\rm
  sub}\approx f_{\rm sub}^{\rm NSE}$ and since $F_{\rm
  h}\approx S_{\rm max}F_{\rm h}^{\rm NSE}$, then $\mathfrak{f}_{\rm
  boost}\approx S_{\rm max}\mathfrak{f}_{\rm boost}^{\rm NSE}$. Thus,
for case i) we have that:
\begin{equation}\label{boost_subs_case_1}
  {b_{\rm sub}}^{(i)}\approx b_{\rm sub}^{\rm NSE}\in(2,60)
\end{equation}
where the numerical values were obtained for the case of no
enhancement in \cite{2010MNRAS.tmp..453Z}.

The case of resonant enhancement is more complex, since the saturation
mass is lower than in the non-resonant case and thus the non-saturated
regime has a more relevant influence on $\mathfrak{f}_{\rm
  boost}$. However, we can show that in general:
\begin{equation}\label{dem}
\mathfrak{f}_{\rm boost}\leq S_{\rm max}\mathfrak{f}_{\rm boost}^{\rm NSE}.
\end{equation}
To prove this we note that since $\mathfrak{f}_{\rm no-boost}\approx
S_{\rm max}\mathfrak{f}_{\rm no-boost}^{\rm NSE}$, thus
$\mathfrak{f}_{\rm boost}=\mathfrak{f}_{\rm no-boost}+[...]=S_{\rm
  max}\mathfrak{f}_{\rm no-boost}^{\rm NSE}+[...]$. Therefore to prove
Eq.~(\ref{dem}) we just need to show that:
\begin{equation}\label{dem2}
\int_{10^{-6}}^{6.89\times10^8}f_{\rm sub}F_{\rm h}\,{\rm d}M_{\rm
  h}\leq S_{\rm max}\int_{10^{-6}}^{6.89\times10^8}f_{\rm sub}^{\rm
  NSE}F_{\rm h}^{\rm NSE}\,{\rm d}M_{\rm h}.
\end{equation}
This is true because each subhalo in a host can be enhanced by $S_{\rm
  max}$ at the most, that means that the total luminosity of all these
subhalos is bounded by $S_{\rm max}$: $f_{\rm sub}L_{\rm h}'\leq
S_{\rm max}(f_{\rm sub}L_{\rm h}')^{\rm NSE}$. Since by definition
$F_{\rm h}\sim {\rm ln}(10)L_{\rm h}'{\rm d}N_{\rm h}/{\rm d}M_{\rm
  h}$, where ${\rm d}N_{\rm h}/{\rm d}M_{\rm h}$ is the number of
halos in the mass range $M_{\rm h}\pm {\rm d}M_{\rm h}$, we have
that: $f_{\rm sub}F_{\rm h}\leq S_{\rm max}(f_{\rm sub}F_{\rm h})^{\rm
  NSE}$, which proves Eq.~(\ref{dem2}).  After doing the calculation
we find that:
\begin{equation}\label{boost_subs_case_2}
  {b_{\rm sub}}^{(ii)}\in(2,42)
\end{equation}
We take the range of values in Eqs.~(\ref{boost_subs_case_1}-\ref{boost_subs_case_2}) 
as extrema reflecting the uncertainties on the extrapolation procedure. They should 
then bracket the true result.

\section{Annihilation in fundamental streams}\label{streams_sec}

\begin{figure}
\center{
\includegraphics[height=8.0cm,width=8.0cm]{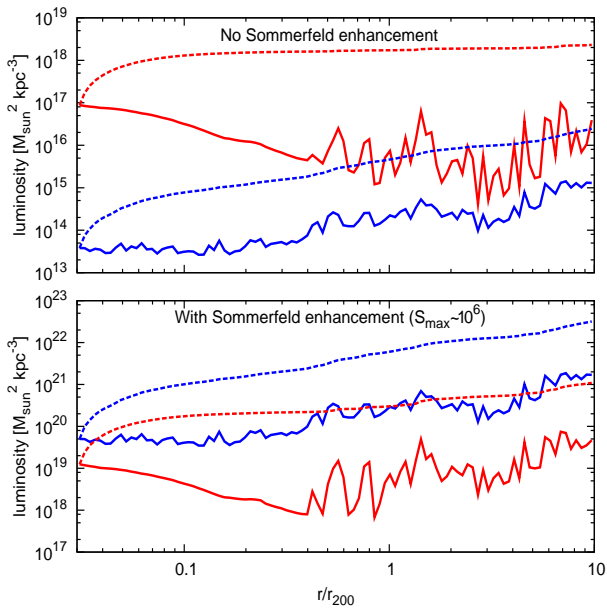}
}
\caption{Differential (solid) and cumulative (dashed) radial profiles of the annihilation luminosity for the
smooth halo component (red) and for the streams component (blue) for the cases with and without Sommerfeld enhancement
in the lower and upper panels respectively. The former, is for the near-resonance case described in section \ref{s_enh} 
that has a saturation value of $S_{\rm max}\sim10^6$.}
\label{fig_streams} 
\end{figure}

To compute the luminosity coming from annihilation in streams, we use the methodology described in
\cite{2009MNRAS.400.2174V,2010arXiv1002.3162V} that integrates the geodesic deviation equation together with the 
N-body equations of motion to follow the evolution of the fine-grained structure of dark matter halos. 
This method was applied to the Milky-Way size objects simulated by the Aquarius project in \cite{2010arXiv1002.3162V}, 
we took their results from one of these objects.

In Fig.~\ref{fig_streams}, we show the spherically averaged radial profiles of the annihilation luminosity for the 
smooth halo component (red), computed using the local mean density, and for the fine-grained intra-stream component (blue).
The solid and dashed lines are for the differential and cumulative profiles respectively. The upper panel shows the case without
Sommerfeld enhancement and the lower one the near-resonance case with a $\sigma_{\rm vel}$-dependent boost factor
($S_{\rm max}\sim10^6$) as described in section \ref{s_enh}. 

Looking at the cumulative distribution in Fig.~\ref{fig_streams}, we see that at the virial radius, $r_{200}$, the ratio of the total 
intra-stream luminosity to the total smooth luminosity is $\sim10^{-3}$ in the case with no enhancement. This ratio increases to 
$\sim20$ once the extreme case of near-resonance enhancement is included. Thus, due to the low velocity dispersion of dark matter 
particles in streams, the annihilation rate in streams dominates over the rate given by the smooth mean density contribution. This
contribution from streams remains nevertheless significantly smaller than the subhalo contribution. Considering subhalos with 
masses down to $10^{-6}$M$_{\odot}$, the ratio of the total subhalo to smooth luminosity for MW-like halos lies in the range: 
$2-2\times10^3$ for the case without Sommerfeld enhancement and $4\times10^3-7\times10^6$ for the near-resonance case (see section
\ref{un_subs}). The subhalo contribution is at least $1000$ times larger than the stream 
contribution when the annihilation cross section is not enhanced by a Sommerfeld mechanism. Once the latter is included, 
it boosts all components (smooth, subhalos and streams) accordingly by a factor which is bounded by $S_{\rm max}$.
Because the subhalo contribution is dominated by the smallest unresolved subhalos and these 
are essentially in the saturated regime, they are boosted by the same amount as the streams, and thus prevail as the dominant
component of the annihilation luminosity in a halo.

\bibliography{lit_PRD}

%\label{lastpage}

\end{document}